\documentclass[journal]{IEEEtran}
\usepackage{cite}

\ifCLASSINFOpdf
   \usepackage[pdftex]{graphicx}
\else
   \usepackage[dvips]{graphicx}
\fi
\usepackage{amsmath}
\usepackage{algorithmic}

\usepackage{array}

\ifCLASSOPTIONcompsoc
 \usepackage[caption=false,font=normalsize,labelfont=sf,textfont=sf]{subfig}
\else
 \usepackage[caption=false,font=footnotesize]{subfig}
\fi

\usepackage{stfloats}
\usepackage{url}

\usepackage[utf8]{inputenc}
\usepackage[T1]{fontenc}
\usepackage{pgf}
\usepackage{xcolor}
\usepackage{physics}
\usepackage{notation}
\usepackage{multirow}
\usepackage{amssymb}
\usepackage{acronym}
\usepackage{psfrag}
\usepackage{soul}
\usepackage{import}
\usepackage[inline]{enumitem}
\usepackage[level=3]{LatexInclusion/messages}
\usepackage{auto-pst-pdf}
\usepackage{tikz}

\acrodef{pmf}[PMF]{probability mass function}
\acrodef{pdf}[PDF]{probability density function}
\acrodefplural{pdf}[PDFs]{probability density functions}
\acrodef{cdf}[CDF]{cumulative distribution function}
\acrodef{mmse}[MMSE]{minimum mean-square error}
\acrodef{da}[DA]{data association}
\acrodef{bp}[BP]{belief propagation}
\acrodef{fg}[FG]{Factor Graph}
\acrodef{nebp}[NEBP]{neural enhanced belief propagation}
\acrodef{mot}[MOT]{multi-object tracking}
\acrodef{jpda}[JPDA]{joint probabilistic data association}
\acrodef{mht}[MHT]{multiple hypothesis tracker}
\acrodef{rfs}[RFS]{random finite sets}
\acrodef{gospa}[GOSPA]{generalized optimal sub-pattern assignment}
\acrodef{rmse}[RMSE]{root mean square error}
\acrodef{mpc}[MPC]{multipath component}
\acrodef{dmc}[DMC]{dense multipath component}
\acrodef{slam}[SLAM]{simultaneous localization and mapping}
\acrodef{2d}[2-D]{two dimensional}
\acrodef{pa}[PA]{physical anchor}
\acrodef{va}[VA]{virtual anchor}
\acrodef{los}[LOS]{line-of-sight}
\acrodef{pf}[PF]{potential feature}
\acrodef{nlos}[NLOS]{non-line-of-sight}
\acrodef{iid}[i.i.d.]{independent and identically distributed}
\acrodef{2d}[2-D]{two-dimensional}
\acrodef{snr}[SNR]{signal-to-noise ratio}
\acrodef{tbd}[TBD]{track-before-detect}
\acrodef{crlb}[CRLB]{Cram\'er-Rao lower bound}

\newcommand{\ist}{\hspace*{.3mm}}
\newcommand{\rmv}{\hspace*{-.3mm}}

\newcommand{\nn}{\nonumber}
\newcommand{\T}{\mathrm{T}}
\newcommand{\CH}{\mathrm{H}}

\definecolor{myblue}{RGB}{79, 129, 189}
\definecolor{myorange}{RGB}{247, 150, 70}

\DeclareMathAlphabet{\mathpzc}{OT1}{pzc}{m}{it}

\begin{document}

\title{Direct Multipath-Based SLAM}

\author{Mingchao~Liang,~\IEEEmembership{Student Member,~IEEE},
        Erik~Leitinger,~\IEEEmembership{Member,~IEEE},
        and~Florian~Meyer,~\IEEEmembership{Member,~IEEE} \vspace*{-5mm}

\thanks{The material presented in this work was supported by the Under Secretary of Defense for Research and Engineering under Air Force Contract No. FA8702-15-D-0001 and by Qualcomm Innovation Fellowship No. 492866.}
\thanks{Mingchao~Liang is with the Department of Electrical and Computer Engineering, University of California San Diego, La Jolla, CA 92093, USA (e-mail: \texttt{m3liang@ucsd.edu}).}
\thanks{Erik~Leitinger is with the Signal Processing and Speech Communication Laboratory, Graz University of Technology, 8010 Graz, Austria (e-mail: \texttt{erik.leitinger@tugraz.at}).}
\thanks{Florian~Meyer is with the Scripps Institution of Oceanography and the Department of Electrical and Computer Engineering, University of California San Diego, La Jolla, CA 92093, USA (e-mail: \texttt{flmeyer@ucsd.edu}).}
          }



\maketitle

\begin{abstract}
In future wireless networks, the availability of information on the position of mobile agents and the propagation environment can enable new services and increase the throughput and robustness of communications. Multipath-based \ac{slam} aims at estimating the position of agents and reflecting features in the environment by exploiting the relationship between the local geometry and \acp{mpc} in received radio signals. Existing multipath-based \ac{slam} methods preprocess received radio signals using a channel estimator. The channel estimator lowers the data rate by extracting a set of dispersion parameters for each \ac{mpc}. These parameters are then used as measurements for \ac{slam}. Bayesian estimation for multipath-based \ac{slam} is facilitated by the lower data rate. However, due to finite resolution capabilities limited by signal bandwidth, channel estimation is prone to errors and \ac{mpc} parameters may be extracted incorrectly and lead to a reduced \ac{slam} performance. We propose a multipath-based \ac{slam} approach that directly uses received radio signals as inputs. A new statistical model that can effectively be represented by a factor graph is introduced. The factor graph is the starting point for the development of an efficient \ac{bp} method for multipath-based \ac{slam} that avoids data preprocessing by a channel estimator. Numerical results based on synthetic and real data in challenging single-input, single-output (SISO) scenarios demonstrate that the proposed method outperforms conventional methods in terms of localization and mapping accuracy\vspace{-1mm}. 
\end{abstract}

\begin{IEEEkeywords}
Simultaneous localization and mapping (SLAM), multipath propagation, Bayesian estimation, belief propagation, factor graphs
\vspace{-4mm}
\end{IEEEkeywords}

%
\IEEEpeerreviewmaketitle

\acresetall
\section{Introduction}
\vspace{-1mm}
\label{sec:introduction}

\Ac{slam} \cite{DisNewClaDurCso:01,MonThrKolWeg:02,CadCarCarLatScaNeiReiLeo:J16,DurBai:J06,MulVoAdaVo:J11,DeuReuDie:J15,GenJosWan:J16,BreAlsYuGla:J17,LeiMeyHlaWitTufWin:J19,WenKulWitWym:J21} aims at estimating the position together with a unknown map of the environment based on noisy measurements provided by sensors hosted on mobile platforms. \ac{slam} is a key signal processing task in a variety of applications, including land robotics, autonomous driving, underwater navigation, and wireless communication. In particular, in future wireless networks, \Ac{slam} can be performed by exploiting multipath propagation of radio signals in urban or indoor scenarios. The resulting knowledge of the agent position and the propagation environment can then be used for, e.g., precise beam tracing or effective resource allocation \cite{RagCezSubSamKoy:J16,DeLBelBerBouDarGuiIsoLohMiaYanBar:J21}\vspace{-3mm}. 


\subsection{Multipath-Based SLAM and State of the Art} \label{sec:introduction_slam}
In wireless networks, localization in scenarios with multiple propagation paths is typically performed using ultra-wideband or mmWave radios. Their relatively high bandwidth makes it possible to resolve \acp{mpc}. 

Early work on radio-based indoor localization aims at mitigating multipath effects \cite{WymMarGifWin:J12}. On the contrary, multipath-based \ac{slam} exploits multiple propagation by inferring the relation between the geometry of the propagation environment and the \acp{mpc} in received radio signals. Existing multipath-based \ac{slam} methods can be categorized as feature-based \ac{slam} \cite{ThrFoxBur:B05,CadCarCar:J16,EbaBerBig:J23}. Here, the map of the environment is represented by an unknown number of individual \textit{map features} with unknown positions. In received radio signals, \acp{mpc} generated by \ac{los} path and specular reflections at flat surfaces often dominate. The map features are referred to as \acp{pa} and \acp{va}, where \acp{pa} are base stations and \acp{va} are the mirror images of \acp{pa} at the reflecting surfaces. The number and position of \acp{va} are unknown. The signals resulting from a transmission from mobile agent to \ac{pa} (or vice versa) provide information on the agent's position, the \ac{pa}, and the \acp{va} associated with it. Consequently, the goal of many multipath-based \ac{slam} approaches is to estimate the positions of the mobile agents together with the number of aforementioned features and their positions. In addition to features related to specular reflections, some multipath-based \ac{slam} methods also aim to estimate the position of point scatterer \cite{GenJosWan:J16} or the parameters of a probability distribution characterizing diffuse multipath \cite{WenKulWitWym:J21}.

Existing multipath-based \ac{slam} methods include methods based on extended Kalman filtering \cite{DisNewClaDurCso:01}, Rao-Blackwellized particle filtering \cite{MonThrKolWeg:02,DurBai:J06,GenJosWan:J16}, sequential estimation of random finite sets \cite{MulVoAdaVo:J11,DeuReuDie:J15}, and \ac{bp} \cite{LeiGreWit:19,LeiVenTeaMey:J23}. All existing approaches perform data processing in two stages. In stage one, a channel estimator is applied to the received radio signal. The channel estimator detects \acp{mpc} and extracts their dispersion parameters, e.g., delay, angle of arrival, angle of departure, or Doppler frequency \cite{WipRao:J07,ShuWanJos:13,GerMecChrXenNan:J16,BadHanThoFle:J17,SanLeiRao:J22,BayPalConZha:22,PalGon:23,GreLeiWitFle:J24}. In stage two, a sequential Bayesian \ac{slam} engine estimates the mobile agent position and the number and position of map features. The extracted dispersion parameters of \acp{mpc} are prone to measurement origin uncertainty. In particular, (i) it is not known which propagation path generated which extracted \ac{mpc}  (ii)  certain propagation paths are missed in the sense that no \ac{mpc} is extracted, and (iii) some \acp{mpc} that do not correspond to any propagation path may be erroneously extracted.

The \ac{slam} engine sequentially estimates agent positions and map features representing propagation paths, addresses measurement-origin uncertainty by performing data association \cite{Kuh:55,BarWilTia:B11,WilLau:J14}, and dynamically introduces new map features in the state space. The use of a channel estimator as a preprocessing stage is widely used as it reduces the data flow and thus leads to SLAM engines that have reduced computational complexity due to simplified models. For example, a key assumption in data association is that each extracted \ac{mpc} can be associated with at most one map feature. On the other hand, the use of a channel estimator can lead to a significant loss of information and thus impede the robustness and accuracy of \ac{slam}. For example, in single-input, single-output (SISO) scenarios, where only delays, i.e., range measurements, can be extracted from the received radio signal, it is often common that multiple \acp{va} have similar ranges to the agent. In such challenging geometries, the channel estimator may not be able to resolve the \ac{mpc} delays and thus extract a single range ``merged'' measurement for multiple \acp{va}. The \ac{slam} engine is then poised to make suboptimum data associations, which can lead to incorrect multipath \ac{slam} estimation results. Since the ability to resolve \ac{mpc} is directly proportional to signal bandwidth, performance degradation related to challenging geometries is exacerbated in low-bandwidth\vspace{-4mm} systems.

\begin{figure*}[!t]
    \centering
    \psfrag{CE1}[c][c][0.80]{\raisebox{-3mm}{Channel}}
    \psfrag{CE2}[c][c][0.80]{\raisebox{-3mm}{parameter}}
    \psfrag{CE3}[c][c][0.80]{\raisebox{0.5mm}{\hspace{-.5mm} estimator}}
    \psfrag{MPC1}[c][c][0.65]{Estimated}  
    \psfrag{MPC2}[c][c][0.65]{MPCs}
    \psfrag{SL1}[c][c][0.80]{Feature-}
    \psfrag{SL2}[c][c][0.80]{based SLAM}
    \psfrag{SL3}[c][c][0.80]{\hspace{2mm}Direct SLAM}
    \psfrag{R1}[c][c][0.80]{Radio signal $\V{z}_k^{(j)}$}
    \psfrag{R2}[c][c][0.80]{$j \in \{1, \dots, J\}$}
    \psfrag{T1}[l][l][0.80]{Previous work}
    \psfrag{T2}[l][l][0.80]{\color{red}{\bf Proposed}}
    \psfrag{RE11}[r][r][0.80]{}
    \psfrag{RE12}[r][r][0.80]{}
    \psfrag{RE13}[r][r][0.80]{Beliefs from}
    \psfrag{RE14}[r][r][0.80]{previous time $k - 1$}
    \psfrag{RE15}[r][r][0.80]{}
    
    \psfrag{RE21}[l][l][0.80]{}
    \psfrag{RE22}[l][l][0.80]{}
    \psfrag{RE23}[l][l][0.80]{Beliefs from}
    \psfrag{RE24}[l][l][0.80]{current time $k$}
    \psfrag{RE25}[l][l][0.80]{}
    \includegraphics[scale=0.75]{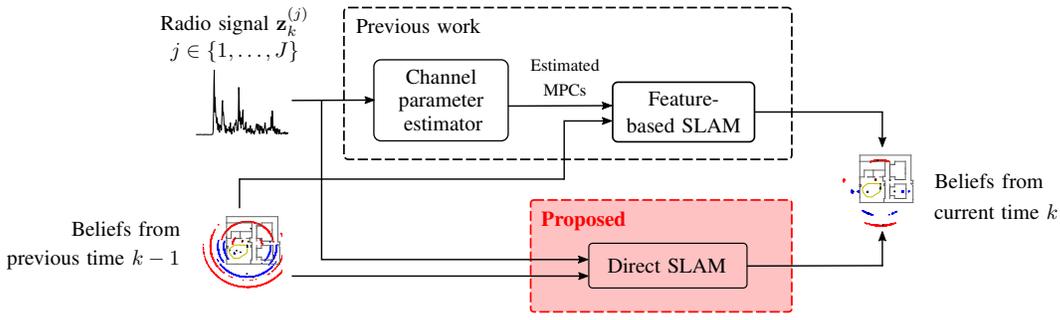}
    \vspace{0mm}
    \caption{Flow diagram of the proposed Direct-SLAM compared to conventional multipath-based SLAM. Discrete time is indexed by $k$. Conventional methods process received radio signals is two stages. In stage one, a channel estimator extracts the parameters of MPCs, e.g., the delays, from $J$ samples of the received radio signal, $\V{z}_{k}^{(j)}, j \in \{1, \dots, J\}$\vspace{-.5mm}. These parameters are then used as the measurements for \ac{slam} in stage two. On the contrary,  Direct-SLAM uses the received radio signal $\V{z}_{k}^{(j)}, j \in \{1, \dots, J\}$ as measurements. Both approachs perform sequential Bayesian estimation, i.e., in addition to the received radio signal, approximate marginal posterior distributions (``beliefs'') from the previous time step $k-1$ are also used as the input at time $k$\vspace{-3mm}.}
    \label{fig:diagram}
    \vspace{-2mm}
\end{figure*}

\subsection{Contributions and Paper Organization} \label{subsec:introduction_contribution}
\vspace{-.5mm}

In this paper, we propose \textit{Direct-SLAM} for multipath-aided localization and mapping. Raw radio signals are directly used as measurements for sequential Bayesian estimation without being preprocessed by a channel estimator to reduce information loss. The proposed method is inspired by \ac{tbd} \cite{VoVoPhaSut:10,RisRosKimWanWil:19,LepRabGla:J16,DavGar:22,LiaKroMey:J23} approaches for multiobject tracking. In particular, it relies on a superpositional measurement model for received radio signals that describes \ac{los} propagation and specular reflections. Other propagation phenomena that can potentially also generate \acp{mpc}, such as point scattering and diffusion, are modeled as interference.  

We introduce a new measurement model for multipath-based \ac{slam} that (i) accurately describes the superposition of specular \acp{mpc}  in received radio signals and a birth model that (ii) introduces new feature hypotheses referred to as \acp{pf}. At each time step, in addition to the state of the mobile agent and the features, the noise variance is also modeled as a random variable. The new statistical model is designed such that it can be efficiently represented by a factor graph \cite{KscFreLoe:01,Loe:04}. It can be seen as a combination of the Swerling 1 model for correlated measurements in \cite{LepRabGla:J16} and the Bernoulli existence model in \cite{LiaKroMey:J23}. The factor graph provides the blueprint for a \ac{bp} method \cite{KscFreLoe:01,YedFreWei:05,KolFri:B09} that can sequentially estimate the number of features and their positions as well as the state of the agent in a scalable and efficient way. \ac{bp}, also known as the sum-product-algorithm \cite{KscFreLoe:01,YedFreWei:05,KolFri:B09}, computes real-value functions called ``messages'' along the edges of the factor graph by exploiting conditional statistical independencies. Most BP messages cannot be calculated in closed form. For an accurate and feasible numerical evaluation, following our previous work in \cite{MeyHliHla:J16,MeyBraWilHla:J17,LiaKroMey:J23}, we represent some of the \ac{bp} messages by random samples ``particles'' and others by means and covariance matrices obtained via moment matching. Contrary to existing \ac{slam} methods, the processing of received radio signals by a channel estimator and explicit data association can be avoided, alleviating performance degradation in challenging \ac{slam} geometries. Basic block diagrams of conventional and proposed methods are shown in Fig.~\ref{fig:diagram}. 

The key contributions of this paper can be summarized as follows.
\begin{itemize}
    \item We introduce a statistical model for multipath-based \ac{slam} that accurately models the data-generating process of received radio signals in SISO\vspace{1.5mm} systems.
    \item We develop a scalable \ac{bp} method for the sequential estimation of agent position and map features based on the factor graph representing the\vspace{1.5mm} new model.
    \item We conduct comprehensive numerical experiments on synthetic and real data and demonstrate state-of-the-art performance.
\end{itemize}
This work advances our preliminary conference publication \cite{LiaLeiMey:C23} by (i) introducing a detailed derivation of the statistical model and the corresponding \ac{bp} methods; (ii) providing a detailed description of the numerical implementation of the resulting algorithm; (iii) developing an alternative computation of \ac{bp} messages that has lower time complexity; (iv) conducting a comprehensive evaluation based on  both synthetic and real data; (v) comparing the agent localization performance of the proposed method with the  \ac{crlb}.

The paper is organized as follows. Section~\ref{sec:model} presents the radio signal model and the new statistical model. Section~\ref{sec:bp} develops the proposed \ac{bp} method for multipath-based \ac{slam}. Section~\ref{sec:particle} introduces a feasible implementation of the \ac{bp} method. Section~\ref{sec:exp} reports the results of the performed numerical study. Finally, Section~\ref{sec:conclusion} concludes the\vspace{-2.5mm} paper.


\section{System Model and Problem Formulation} \label{sec:model}

In what follows, we introduce the proposed system model and discuss the problem formulation of the considered multipath-based \ac{slam}\vspace{-3.8mm} problem.

\subsection{Radio Signal and Specular MPCs} \label{subsec:radio_signal}
\vspace{-.5mm}

Multipath effects in radio channels are directly related to the geometry of the propagation environment \cite{Mei:14,GenJosWan:J16,LeiGreWit:19}. In particular, the parameters of specular \acp{mpc}, e.g., delays, can be described by the position of the mobile agent and the positions of \acp{pa} and \acp{va}. In what follows, we assume the mobile agent acts as a transmitter, and the \acp{pa} act as receivers. Note, however, that the proposed model can be easily reformulated for the case where the roles of \acp{pa} and mobile agent are swapped.

\begin{figure}[!h]
	\centering
	\psfrag{A1}[c][c][0.80]{\raisebox{-0mm}{$\hspace{0mm}\V{p}_k$}}
	\psfrag{B1}[c][c][0.80]{\raisebox{1mm}{$\hspace{-0mm}\V{p}_{2}^{(1)}$}}
    \psfrag{B2}[c][c][0.80]{\raisebox{3mm}{$\hspace{-0mm}\V{p}_{3}^{(1)}$}}
	\psfrag{C1}[l][l][0.80]{\raisebox{1mm}{$\hspace{-.0mm} \V{p}_{1}^{(1)}$}}
	\psfrag{W1}[l][l][0.80]{Reflecting Surface}
	\psfrag{W2}[l][l][0.80]{LOS Path}
	\psfrag{W3}[l][l][0.80]{Reflected Path}
	\psfrag{W4}[l][l][0.80]{Agent}
	\psfrag{W5}[l][l][0.80]{VA}
	\psfrag{W6}[l][l][0.80]{PA}
    \psfrag{S1}[l][l][0.80]{Surface 2}
	\psfrag{S2}[l][l][0.80]{Surface 1}
	\includegraphics[scale=0.85]{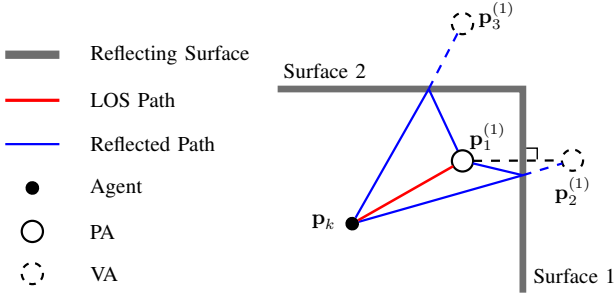}
	\vspace{.5mm}
	\caption{Scenario with one mobile agent, one physical anchor (PA), and two virtual anchors (VAs). The radio signal transmitted by the agent at position $\V{p}_k$ propagates to the receiver at position $\V{p}_1^{(1)}$ on the LOS path and on the reflected paths. The path length of the reflected paths are equal to the length of the paths from the agent at position $\V{p}_k$ to the VAs at position $\V{p}_2^{(1)}$ and $\V{p}_3^{(1)}$, respectively\vspace{-1mm}. }
	\label{fig:va_example}
\end{figure}

\ac{va} positions are the mirror images of \ac{pa} positions at flat reflecting surfaces and encode the geometry of the environment. In particular, the length of the path from the mobile agent to the \ac{va} is equal to the length of the path from the mobile agent to the reflecting surface to the \ac{pa}. This principle is illustrated in\vspace{-1.5mm} Fig.~\ref{fig:va_example}. 


For the considered multipath-based \ac{slam} problem, we consider a single mobile agent with an unknown position. At discrete time $k$, the position of the agent is denoted as  $\V{p}_k$. In addition, there are $J$ \acp{pa} with known positions $\V{p}_{1}^{(j)}, j \in \{1, \dots, J\}$. At time step $k$, each \ac{pa} $j$ has $L_{k}^{(j)} - 1$ associated \acp{va}, with unknown positions $\V{p}_{l}^{(j)}, l \in \{2, \dots, L_{k}^{(j)}\}$. \acp{pa} and \acp{va}, here referred to as features, provide a partial characterization of the propagation environment. While \ac{pa} positions are known, due to the unavailability of a floor plan, \ac{va} positions are unknown. The fact that the absolute position of certain features in the environment, i.e., the \acp{pa}, is known, is a major difference of multipath-aided SLAM compared to traditional robotic SLAM.

The mobile agent transmits a signal at center frequency $f_\mathrm{c}$ and bandwidth $B$. In the frequency domain, the transmitted signal in baseband is $S(f) \rmv\in\rmv \mathbb{C}$. At each discrete time step $k$, $M$ samples of the received signal waveform are recorded by the \acp{pa}. In the frequency domain, the resulting $M = B/\Delta + 1$ samples have a frequency spacing of $\Delta$\footnote{The corresponding unambiguous maximum observation distance is given by $d_\mathrm{max} = c/\Delta$.}\hspace{-.5mm}. Let $\V{h}(t)$ denote the sampled transmitted signal, delayed by time $t$, in baseband and frequency domain\vspace{0mm}, i.e.,
\begin{align}
    \V{h}(t) &= \big[S \big( -(M-1)/2 \ist \Delta \big) \ist\mathrm{e}^{j2\pi (M-1)/2 \ist \Delta t} \nn\\[1mm] &\hspace{5mm} \cdots \hspace{1mm} S \big( (M-1)/2 \ist \Delta \big) \ist\mathrm{e}^{-j2\pi (M-1)/2 \ist \Delta t}\big]^\T \rmv\rmv. \label{eq:ref_signal}
\end{align} 
The measurement model for the sampled received signal at PA $j$, i.e.,  $\V{z}^{(j)}_{k} = \big[z_k^{(j)}(-(M-1)/2 \ist \Delta) \hspace{1mm} \cdots \hspace{1mm} z_k^{(j)} ( (M-1)/2 \ist \Delta )\big]^\T = \big[z_{k,1}^{(j)} \hspace{1mm} \cdots \hspace{1mm} z_{k,M}^{(j)} \big]^\T\rmv\rmv\rmv$, can now be expressed as \cite{LeiGreFleWit:20,Ric:05}\vspace{-3.5mm}
\begin{align}
    \V{z}^{(j)}_{k} = \sum_{l = 1}^{L_k^{(j)}} \varrho_{k, l}^{(j)} \ist\ist \V{h}(t_{k, l}^{(j)}) + \V{\epsilon}^{(j)}_{k} \label{eq:radio_signal}
\end{align}
where $t_{k, l}^{(j)} = \Vert \V{p}_k - \V{p}_{l}^{(j)} \Vert / c$ is the delay related to the $l$-th \ac{mpc}, $\varrho_{k, l}^{(j)} \in \mathbb{C}$ is the complex amplitude of propagation path $j$, and $c$ is the speed of light. For example, a simple physical model for the complex amplitude, $\varrho_{k, l}^{(j)} \in \mathbb{C}$, is given by\vspace{-.3mm} 
\begin{align}
	\varrho_{k, l}^{(j)} = a^{(j)}_{l,k}\ist \mathrm{e}^{-j2\pi f_\mathrm{c}t_{k, l}^{(j)}} \frac{c}{4\pi f_\mathrm{c}\Vert \V{p}_k - \V{p}_{l}^{(j)} \Vert}\label{eq:complex_amplitudes}
\end{align}
where $a^{(j)}_{l,k} \in \mathbb{C}$ is an unknown cumulative reflection coefficient resulting from all interactions of the radio signal with flat surfaces\footnote{The signal model in \eqref{eq:radio_signal} implies uniform radiation patterns for transmit and receive antennas.}. Furthermore, $\V{\epsilon}^{(j)}_{k} =  \big[ \epsilon_{k}^{(j)} \big( \frac{-(M - 1)}{2}  \Delta \big) $ $ \cdots \hspace{1mm} \epsilon_{k}^{(j)} ( \frac{M - 1}{2} \Delta \ist \big) \big]^\T\rmv\rmv\rmv \in \mathbb{C}^M$ is the noise vector.

\subsection{State Vectors and Measurement Model} \label{subsec:meas_model}

The state of the mobile agent is denoted as $\V{x}_{k}$. It includes the agent's position $\V{p}_k$ and possibly further motion-related parameters. The number of features $L_{k}^{(j)}, j \in \{1, \dots, J\}$ associated with each \ac{pa} depends on the position of the agent and is generally unknown and time-varying. For the estimate of the unknown $L_{k}^{(j)}\rmv\rmv$ at each time $k$, so called \acp{pf} indexed by $n \in \{1, \dots, N_{k}^{(j)}\}$ are introduced \cite{MeyBraWilHla:J17,LeiMeyHlaWitTufWin:J19}. Here $N_{k}^{(j)}$ is the maximum possible number of features associated with the $j$-th \ac{pa} at time step $k$. The existence of each \ac{pf} $j \in \{1, \dots, J\}$ is modeled by a binary random variable $r_{k, n}^{(j)} \in \{0, 1\}, n \in \{1, \dots, N_{k}^{(j)}\}$. A \ac{pf} exists, i.e., corresponds to a physical feature in the environment contributing to the signal received at \ac{pa} $j$, if and only if $r_{k, n}^{(j)} \rmv=\rmv 1$.

The state of \acp{pf} is defined as $\V{y}_{k, n}^{(j)} = [\V{\phi}_{k, n}^{(j) \T} \hspace{1mm} r_{k, n}^{(j)}]^\T$ where $\V{\phi}_{k, n}^{(j)} = [\V{p}_{k, n}^{(j) \T} \hspace{1mm} \gamma_{k, n}^{(j)}]^\T$ includes the position $\V{p}_{k, n}^{(j)}$ and the intensity $\gamma_{k, n}^{(j)}$. The intensity represents the signal strength of the propagation path represented by the \ac{pf}. Since each \ac{pa} is also modeled as a \ac{pf}, we have $N_k^{(j)} \ge 1$. In particular, we always use the first index, $n = 1$, for \acp{pf} representing \acp{pa}. For future reference, we establish the notation $\V{y}^{(j)}_{k} = \big[\V{y}^{(j) \T}_{k, 1} \cdots \V{y}^{(j) \T}_{k, N_k^{(j)}}\big]^\T\rmv\rmv\rmv$.


Based on the \ac{pf} representation, the measurement model in \eqref{eq:radio_signal}, can\vspace{-.5mm} also be expressed as
\begin{align}
    \V{z}^{(j)}_{k} = \sum^{N_k^{(j)}}_{n = 1} \ist r_{k, n}^{(j)} \ist\ist \rho_{k, n}^{(j)} \ist \V{h}_{k, n}^{(j)} + \V{\epsilon}^{(j)}_{k} \label{eq:meas_model}\\[-6mm]\nn
\end{align}
where $\V{h}_{k, n}^{(j)} \triangleq \V{h}(\tau_{k, n}^{(j)}) \in \mathbb{C}^M$ denotes\vspace{-.3mm} the known contribution vector related to \ac{pf} with indexes $j \in \{1, \dots, J\}$  and $n \in \{1, \dots, N_{k}^{(j)}\}$, i.e., $\tau_{k, n}^{(j)} = \Vert \V{p}_k - \V{p}_{k, n}^{(j)} \Vert / c$. Following \cite{ShuWanJos:13,WipRao:J07,GerMecChrXenNan:J16,BadHanThoFle:J17,SanLeiRao:J22}, we here assume that the complex\vspace{0mm} amplitudes, $\rho_{k, n}^{(j)}$, are random and unknown. In particular, random amplitudes are modeled by zero-mean complex Gaussian random variables with variance $\gamma^{(j)}_{k,n}$, i.e., \vspace{0mm} $\rho_{k, n}^{(j)} \sim \mathcal{CN}\big(\rho_{k, n}^{(j)}; 0,\gamma^{(j)}_{k,n}\big)$. 
This amplitude model is referred to as \textit{Swerling} 1 \cite{Sko:B00, LepRabGla:J16}. The complex amplitudes $\rho_{k, n}^{(j)}$ are independent across $k, n$, and $j$. The noise $\V{\epsilon}^{(j)}_{k}\rmv\rmv$ represents different noise sources including \ac{dmc}. For the sake of simplcity $\V{\epsilon}^{(j)}_{k}\rmv\rmv$ is assumed distributed according to $\mathcal{CN}\big(\V{\epsilon}^{(j)}_{k}; \V{0}, \eta_{k}^{(j)} \M{I}_M \big)$ and independent across $k$ and all $j$, as well as independent of all complex amplitudes.

As a result of this model, conditioned on the agent state $\V{x}_k$, the joint \ac{pf} state $\V{y}_{k}^{(j)}$, and the noise variance $\eta_{k}^{(j)}$\vspace{-.5mm}, the measurement $\V{z}^{(j)}_{k}$ is also zero-mean complex Gaussian, i.e.,
\begin{equation}
    f(\V{z}_{k}^{(j)} | \V{x}_k, \V{y}_{k}^{(j)}, \eta_{k}^{(j)}) = \mathcal{CN}(\V{z}_{k}^{(j)}; \V{0}, \M{C}_{k}^{(j)}) \label{eq:likelihood}
\end{equation}
with covariance $\M{C}^{(j)}_{k} = \eta_{k}^{(j)} \M{I}_M + \sum^{N_k^{(j)}}_{n = 1} r_{k, n}^{(j)} \gamma_{k, n}^{(j)}$ $\V{h}^{(j)}_{k, n} \V{h}^{(j) \CH}_{k, n}$.
Finally, we\vspace{-0.5mm} have $f(\V{z}_{k} | \V{x}_k, \V{y}_{k}, \V{\eta}_{k}) = \prod_{j = 1}^{J}$ $f(\V{z}_{k}^{(j)} |$ $\V{x}_k, \V{y}_{k}^{(j)}, \eta_{k}^{(j)})$, where\vspace{0mm} we introduced $\V{z}_{k} = \big[\V{z}_{k}^{(1) \T} \cdots \V{z}_{k}^{(J) \T}\big]^\T\rmv\rmv\rmv$, $\V{y}_{k} = \big[\V{y}_{k}^{(1) \T} \cdots \V{y}_{k}^{(J) \T}\big]^\T\rmv\rmv\rmv$, and $\V{\eta}_{k} = \big[\eta_{k}^{(1)} \cdots \eta_{k}^{(J)}\vspace{-2mm}\big]^\T$\vspace{0mm}.

\subsection{State-Transition Models and Prior PDFs} \label{subsec:state_transition_model}
The dynamics of the agent state $\V{x}_{k}$ are described statistically by a first-order Markov model. Its state transition \ac{pdf} is denoted as $f(\V{x}_{k} | \V{x}_{k - 1})$, which can, e.g., be the result of a constant velocity model \cite[Ch. 4]{ShaKirLi:B02}. The dynamics of the noise variance is described by a \ac{pdf} $f(\eta_{k}^{(j)} | \eta_{k - 1}^{(j)})$, which can, e.g., be a Gamma distribution. Each \ac{pf} state $\V{y}_{k, n}^{(j)}$  with $n \in \{1, \dots, N_{k - 1}^{(j)}\}$ and $j \in \{1, \dots, J\}$ is also assumed to evolve independently following a first-order\vspace{-.5mm} Markov model.
The individual state-transition \ac{pdf} of the \ac{pf} with indexes $n \in \{1, \dots, N_{k-1}^{(j)}\}$ and $j \in \{1,\dots,$ $J\}$, is denote by $f(\V{y}_{k, n}^{(j)} | \V{y}_{k - 1, n}^{(j)}) \rmv\rmv = f(\V{\phi}_{k, n}^{(j)}, r_{k, n}^{(j)} | \V{\phi}_{k - 1, n}^{(j)}, r_{k - 1, n}^{(j)})\vspace{.3mm}$. 
 If \ac{pf} $(j, n)$ does not exist at time step $k - 1$, i.e, $r_{k - 1, n}^{(j)} = 0$, then it does not\vspace{-.5mm} exist at time step $k$ either\vspace{0mm}. The state-transition \ac{pdf} for $r_{k - 1, n}^{(j)} = 0$ is thus given\vspace{-1.5mm} by
\begin{equation}
    f(\V{\phi}_{k, n}^{(j)}, r_{k, n}^{(j)} | \V{\phi}_{k - 1, n}^{(j)}, 0) = \begin{cases}
        f_{\mathrm{D}}(\V{\phi}_{k, n}^{(j)}), & r_{k, n}^{(j)} = 0 \\
        0, & r_{k, n}^{(j)} = 1
    \end{cases} \label{eq:state_transition_pf1}
\end{equation}
where $f_{\mathrm{D}}(\V{\phi}_{k, n}^{(j)})$ is an arbitrary ``dummy'' \ac{pdf}. However, if the \ac{pf} exists at time step $k - 1$, i.e, $r_{k - 1, n}^{(j)} = 1$, then it continues to exist at time step $k$ with survival probability $0 < p_{\mathrm{s}} \le 1$. If it still exists at time $k$, then $\V{\phi}_{k, n}^{(j)}$ is distributed according to $f(\V{\phi}_{k, n}^{(j)} | \V{\phi}_{k - 1, n}^{(j)})$, which can be, e.g., a random walk model. For $r_{k - 1, n}^{(j)} = 0$, the state-transition \ac{pdf} thus\vspace{-.5mm} reads
\begin{align}
    f(\V{\phi}_{k, n}^{(j)}, r_{k, n}^{(j)} | \V{\phi}_{k - 1, n}^{(j)}, 1) = \begin{cases}
        (1 - p_{\mathrm{s}}) f_{\mathrm{D}}(\V{\phi}_{k, n}^{(j)}), & r_{k, n}^{(j)} = 0 \\
        p_{\mathrm{s}} f(\V{\phi}_{k, n}^{(j)} | \V{\phi}_{k - 1, n}^{(j)}), & r_{k, n}^{(j)} = 1.
    \end{cases} \nonumber \\
     \label{eq:state_transition_pf2}\\[-7.5mm]
     \nonumber
\end{align}

For all the state-transition models, we make the following assumptions. Conditioned on $\V{y}_{k - 1, n}^{(j)}, j \in \{1, \dots, J\}, n \in \{1, \dots, N_{k - 1}^{(j)} \}$, the current \ac{pf} state $\V{y}_{k, n}^{(j)}$ is independent of the current and previous agent states, the current and previous measurement noise variances, and current and previous states of the other \acp{pf}. This implies that the joint state transition function of \ac{pf} states factorizes across $n$ and $j$ (see \cite{LiaLeiMey:J24SM} for details). Similarly, we assume that, conditioned on $\V{x}_{k - 1}$, $\V{x}_k$ is independent of current and previous states of all \acp{pf} and all current and previous noise variances. The noise variance $\eta_{k}^{(j)}$ is also assumed to be independent of current and previous states of the agent and all \acp{pf}, conditioned on $\eta_{k - 1}^{(j)}$.



The birth of newly appearing features associated with \ac{pa} $j$ and time $k$ is modeled by a point Poisson process with mean $\mu_{\mathrm{B}}^{(j)}$ and spatial \ac{pdf} $f_{\mathrm{B}}^{(j)}(\V{\phi}| \V{x}_{k})$. Assuming that (i) the number of newly appearing features is significantly smaller than the number of measurements and (ii) newly appearing features are well separated, we introduce $M$ new \acp{pf} at each time step, i.e., $N_{k}^{(j)} = N_{k - 1}^{(j)} + M$ \cite{LiaKroMey:J23}. The initial \ac{pdf} of each new \ac{pf}, conditioned on $\V{x}_{k}$, can thus be written\vspace{-.8mm} as 
\begin{equation}
    f_{\mathrm{B}, n}^{(j)}(\V{\phi} | \V{x}_{k}) \propto \begin{cases}
        f_{\mathrm{B}}^{(j)} (\V{\phi} | \V{x}_{k}), & \V{p} \in \Set{P}_m(\V{x}_{k}) \\
        0, & \V{p} \not\in \Set{P}_m(\V{x}_{k})
    \end{cases} \nn \vspace{.5mm}
\end{equation}
where $\propto$ indicates equality up to a multiplicative constant and the region $\Set{P}_m(\V{x}_{k})$ that is occupied by index $m = n - N_{k - 1}^{(j)} \in \{1, \dots, M\}$, is given by
\begin{equation}
    \Set{P}_m(\V{x}_{k}) = \{\V{p} \hspace{1mm} | \hspace{1mm} (m - 1) \cdot T_s \le \Vert \V{p} - \V{p}_k \Vert / c \le m \cdot T_s \}. \label{eq:birth_pos_set}
\end{equation}

To define the existence probability of new \acp{pf}, we first note that the number of new features in cell $\Set{P}_m(\V{x}_k)$ is also Poisson distributed with mean $\mu_{\mathrm{B}, n}^{(j)} = \mu_{\mathrm{B}}^{(j)} \int \hspace{-1mm} \int_{\Set{P}_m(\V{x}_k)} f_{\mathrm{B}}^{(j)} (\V{\phi} | \V{x}_{k}) \ist \mathrm{d} \V{p} \hspace{.3mm} \mathrm{d} \gamma$, and the probabilities that each cell has zero and one feature are $\mathrm{e}^{-\mu_{\mathrm{B}, n}^{(j)}}$ and $\mu_{\mathrm{B}, n}^{(j)} \mathrm{e}^{-\mu_{\mathrm{B}, n}^{(j)}}$, respectively. Based on the assumption that there is at most one new feature in $\Set{P}_m(\V{x}_k)$, we define the existence probability of the corresponding \ac{pf}{\vspace{0mm} as 
\begin{equation}
    p_{\mathrm{B}, n}^{(j)} = \frac{\mu_{\mathrm{B}, n}^{(j)} \mathrm{e}^{-\mu_{\mathrm{B}, n}^{(j)}} }{\mu_{\mathrm{B}, n}^{(j)} \mathrm{e}^{-\mu_{\mathrm{B}, n}^{(j)}} + \mathrm{e}^{-\mu_{\mathrm{B}, n}^{(j)}}} = \frac{\mu_{\mathrm{B}, n}^{(j)}} {\mu_{\mathrm{B}, n}^{(j)} + 1}. \nn
\end{equation}
As a result, the  \ac{pdf} for individual new \acp{pf}\vspace{.5mm} reads
\begin{align}
    f(\V{y}_{k, n}^{(j)} | \V{x}_{k}) = \begin{cases}
    (1 - p_{\mathrm{B}, n}^{(j)}) \ist f_{\mathrm{D}}(\V{\phi}_{k, n}^{(j)}), & r_{k, n}^{(j)} = 0 \\[1mm]
    p_{\mathrm{B}, n}^{(j)} \ist f_{\mathrm{B}, n}^{(j)} (\V{\phi}_{k, n}^{(j)} | \V{x}_{k}), & r_{k, n}^{(j)} = 1 \ist.\\
    \end{cases} \label{eq:birth_pf} \\[-5.5mm]
    \nn
\end{align}
We assume that, conditioned on $\V{x}_k$, the new \ac{pf} state $\V{y}_{k, n}^{(j)}, j \in \{1, \dots, J\}, n \in \{N_{k - 1}^{(j)} + 1, \dots, N_{k}^{(j)} \}$ is independent of all previous agent states, all current and previous states of all other \ac{pf}, and the current and previous noise covariance matrices. The joint initial \ac{pdf} of new \acp{pf} at time $k$ can thus be obtained as the product of the individual initial \acp{pdf} of new \acp{pf} (see \cite{LiaLeiMey:J24SM} for details). Note that based on the binary random variables $r_{k, n}^{(j)} \in \{0, 1\}$ (cf. Section~\ref{subsec:meas_model}) that indicate feature existence and the Poisson point process used to describe the appearance of new \acp{pf} discussed above, we can represent dynamic \ac{slam} environments with features that appear/disappear due to limited visibility, e.g., blockage of propagation paths and changes in the environment\vspace{2mm}.


At time $k=0$, the prior distributions $f(\V{x}_0)$, $f(\eta_{0}^{(j)})$, $j \in \{1, \dots, J\}$, and $f(\V{y}^{(j)}_{0, n})$, $n \in \{1, \dots, N_{0}^{(j)}\}$, $j \in \{1, \dots, J\}$ are assumed known. The random variables $\V{x}_0$, $\eta_{0}^{(j)}$, $j \in \{1, \dots, J\}$, and $\V{y}^{(j)}_{0, n}$, $n \in \{1, \dots, N_{0}^{(j)}\}$, $j \in \{1, \dots, J\}$ are all independent of each other\vspace{-3mm}. 



\subsection{Declaration and Estimation} \label{subsec:declaration_estimation}
\vspace{-.5mm}

At each time step $k$, given all the measurements $\V{z}_{1 : k} = [\V{z}_{1}^\T \cdots \V{z}_k^\T]^\T\rmv\rmv$, the goal of multipath-based \ac{slam} is to declare the existence of \acp{pf} as well as estimate the states $\V{\phi}_{k, n}^{(j)}$ of \ac{pf} that have been declared to exist, the agent state $\V{x}_{k}$, and the noise variance $\eta_{k}^{(j)}$. In our Bayesian setting, the problem comprises the computation of the marginal \acp{pdf} $f(r_{k, n}^{(j)} | \V{z}_{1 : k})$, $f(\V{\phi}_{k, n}^{(j)} | r_{k, n}^{(j)} = 1, \V{z}_{1 : k})$, $f(\V{x}_{k} | \V{z}_{1 : k})$, and $f(\eta_{k}^{(j)} | \V{z}_{1 : k})$. We refer to $f(r_{k, n}^{(j)} = 1 | \V{z}_{1 : k})$, $n \in \{1, \dots, N_{k}^{(j)}\}$ as the existence probability of \acp{pf}. A \ac{pf} is declared to exist, if its existence probability is larger than a certain fixed threshold $T_{\mathrm{dec}}$. 

Based on these marginal \acp{pdf}, the \ac{mmse} estimate of the agent states, \acp{pf} declared to exist, and noise variance at time $k$, can be obtained as\vspace{-2.5mm}
\begin{align}
    \hat{\V{x}}_{k, \mathrm{MMSE}} &= \int \V{x}_{k} f(\V{x}_{k} | \V{z}_{1 : k}) \hspace{1mm} \mathrm{d} \V{x}_{k} \label{eq:mmse_x} \\[.5mm]
    \hat{\V{\phi}}_{k, n, \mathrm{MMSE}}^{(j)} &= \int \V{\phi}_{k, n}^{(j)} f(\V{\phi}_{k, n}^{(j)} | r_{k, n}^{(j)} = 1, \V{z}_{1 : k}) \hspace{1mm} \mathrm{d} \V{\phi}_{k, n}^{(j)} \label{eq:mmse_y} \\[.5mm]
    \hat{\eta}_{k, \mathrm{MMSE}}^{(j)} &= \int \eta_{k}^{(j)} f(\eta_{k}^{(j)} | \V{z}_{1 : k}) \hspace{1mm} \mathrm{d} \eta_{k}^{(j)}. \label{eq:mmse_epsilon} \\[-5.5mm]
    \nn
\end{align}
Note that $f(r_{k, n}^{(j)} = 1 | \V{z}_{1 : k})$ and $f(\V{\phi}_{k, n}^{(j)} | r_{k, n}^{(j)} = 1, \V{z}_{1 : k})$\vspace{-.6mm} can be computed from the marginal posterior \ac{pdf} $f(\V{y}_{k, n}^{(j)} | \V{z}_{1 : k}) = f(\V{\phi}_{k, n}^{(j)}, r_{k, n}^{(j)} | \V{z}_{1 : k})$. 

Following the statistical model and assumptions in Sections~\ref{subsec:radio_signal} to \ref{subsec:state_transition_model}, the joint posterior \ac{pdf} $f(\V{x}_{0 : k}, \V{y}_{0 : k}, \V{\eta}_{0 : k} | \V{z}_{1 : k})$ can be factorized\vspace{1mm} as
\begin{align}
    f(\V{x}_{0 : k}, \V{y}_{0 : k}, \V{\eta}_{0 : k} | \V{z}_{1 : k}) \hspace{-35mm}& \nn \\[.3mm]
    &\propto f(\V{x}_0) \bigg( \prod^J_{j = 1} f(\eta_{0}^{(j)}) \prod_{n = 1}^{N^{(j)}_0} f(\V{y}^{(j)}_{0, n} ) \bigg) \prod_{k^\prime = 1}^{k} f(\V{x}_{k'} | \V{x}_{k' - 1}) \nn \\[.3mm]
    & \times \prod^J_{j = 1} \bigg( \prod_{n = 1}^{N^{(j)}_{k^\prime - 1}} f(\V{y}^{(j)}_{k^\prime, n} | \V{y}^{(j)}_{k^\prime - 1, n}) \bigg) \bigg(  \prod_{n = N^{(j)}_{k^\prime - 1} + 1}^{N^{(j)}_{k^\prime}} \rmv\rmv\rmv\rmv f(\V{y}^{(j)}_{k^\prime\rmv, n} | \V{x}_{k'}) \rmv \bigg)  \nn \\[1.5mm] 
    & \times f(\eta_{k'}^{(j)} | \eta_{k' - 1}^{(j)}) \ist f(\V{z}^{(j)}_{k'} | \V{x}_{k'}, \V{y}^{(j)}_{k'}, \eta_{k'}^{(j)})  \label{eq:factorization}\\[-4mm]
   \nn
\end{align}
where we introduced $\V{x}_{0 : k} \triangleq [\V{x}_0^\T \cdots \V{x}_{k}^\T]^\T$, $\V{y}_{0 : k} \triangleq [\V{y}_0^\T \cdots \V{y}_{k}^\T]^\T\rmv\rmv$, and $\V{\eta}_{0 : k} \triangleq [\V{\eta}_{0}^\T \cdots \V{\eta}_{k}^\T]^\T$. A detailed derivation is provided in \cite{LiaLeiMey:J24SM}. A single time step of the factor graph corresponding to the factorization in \eqref{eq:factorization}, is shown in Fig.~\ref{fig:factor_graph}. This factorization facilitates the development of an efficient \ac{bp} method for the computation of approximate marginal posterior \acp{pdf} $\tilde{f}(\V{y}_{k, n}^{(j)}) \rmv\approx\rmv f(\V{y}_{k, n}^{(j)} | \V{z}_{1 : k})$, $\tilde{f}(\V{x}_{k}) \rmv\approx\rmv f(\V{x}_{k} | \V{z}_{1 : k})$, and $\tilde{f}(\eta_{k}^{(j)})$ $\approx f(\eta_{k}^{(j)} | \V{z}_{1 : k})$, typically referred to as\vspace{-3mm} beliefs.

%

\begin{figure*}[!htbp]
\vspace{-2mm}
    \centering
    \psfrag{da1}[c][c][0.7]{\raisebox{-2mm}{\hspace{.8mm}$\V{y}_{1}$}}
    \psfrag{daI}[c][c][0.7]{\raisebox{-2.5mm}{\hspace{.3mm}$\V{y}_{\underline{N}}$}}
    \psfrag{db1}[c][c][0.65]{\raisebox{-2mm}{\hspace{.2mm}$\V{y}_{\scriptscriptstyle \underline{N} + 1}$}}
    \psfrag{dbJ}[c][c][0.7]{\raisebox{1mm}{$\V{y}_{N}$}}
    \psfrag{q1}[c][c][0.7]{\raisebox{-1mm}{$f_{1}$}}
    \psfrag{qI}[c][c][0.7]{\raisebox{-2mm}{$f_{\underline{N}}$}}
    \psfrag{v1}[c][c][0.65]{\raisebox{-2.3mm}{\hspace{.15mm}$f_{\scriptscriptstyle \underline{N} + 1}$}}
    \psfrag{vJ}[c][c][0.7]{\raisebox{-1mm}{\hspace{.3mm}$f_{N}$}}
    \psfrag{g1}[c][c][0.7]{\raisebox{-1mm}{\hspace{0mm}$f_{\V{z}}$}}
    \psfrag{ma11}[r][r][0.7]{\color{blue}{\raisebox{0mm}{$\alpha_{1}$}}}
    \psfrag{ma12}[r][r][0.7]{\color{blue}{\raisebox{-2mm}{$\alpha_{1}$}}}
    \psfrag{maJ1}[l][l][0.7]{\color{blue}{$\alpha_{N}$}}
    \psfrag{maJ2}[l][l][0.7]{\color{blue}{$\alpha_{N}$}}
    \psfrag{mkJ}[c][c][0.7]{\raisebox{-0mm}{\color{blue}{$\kappa_{N}$}}}
    \psfrag{mk1}[c][c][0.7]{\color{blue}{\raisebox{-0mm}{$\kappa_{1}$}}}
    \psfrag{mx11}[r][r][0.7]{\color{blue}{$\beta$}}
    \psfrag{mx12}[r][r][0.7]{\color{blue}{$\beta$}}
    \psfrag{mx13}[l][l][0.7]{\color{blue}{$\beta$}}
    \psfrag{mx2}[l][l][0.7]{\color{blue}{$\iota$}}
    \psfrag{dx}[c][c][0.7]{$\V{x}$}
    \psfrag{qx}[c][c][0.7]{$f$}
    \psfrag{de}[c][c][0.7]{\raisebox{-2mm}{$\eta$}}
    \psfrag{qe}[c][c][0.7]{$f_{\eta}$}
    \psfrag{me1}[r][r][0.7]{\color{blue}{$\xi$}}
    \psfrag{me2}[r][r][0.7]{\color{blue}{$\xi$}}
    \psfrag{mn2}[l][l][0.7]{\color{blue}{$\nu$}}
    \psfrag{pa1}[c][c][1]{PA $1$}
    \psfrag{paJ}[c][c][1]{PA $J$}
    \includegraphics[width=0.98\linewidth]{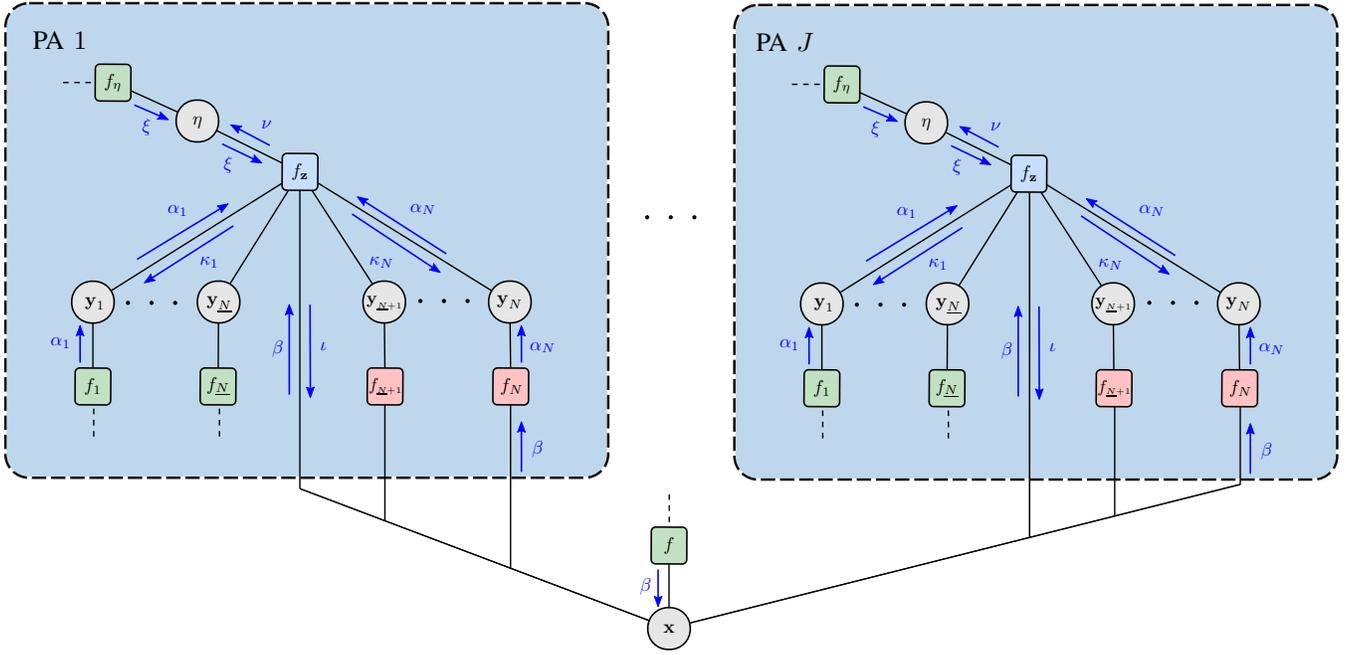}
    \vspace{-2.5mm}
    \caption{Factor graph representing the joint posterior distribution in \eqref{eq:factorization} and the \ac{bp} messages for a single time step $k$. Each box with light blue background indicates a different \acp{pa}. \ac{bp} messages are shown as blue arrows. The time index $k$ is omitted. The following shorthand notation is used: $\underline{N} = N_{k - 1}^{(j)}$, $N = N_k^{(j)}$, $\V{x} = \V{x}_k$, $\V{y}_{n} = \V{y}_{k, n}^{(j)}$, $\eta = \eta_{k}^{(j)}$, $f = f(\V{x}_k | \V{x}_{k - 1})$, $f_{\eta} = f(\eta_{k}^{(j)} | \eta_{k - 1}^{(j)})$, $f_{\V{z}} = f(\V{z}_k^{(j)} | \V{x}_k, \V{y}_{k}^{(j)}, \eta_{k}^{(j)})$, $f_n = f(\V{y}_{k, n}^{(j)} | \V{y}_{k - 1, n}^{(j)})$ for $n \in \{1, \dots, N_{k - 1}^{(j)}\}$, and $f_n = f(\V{y}_{k, n}^{(j)} | \V{x}_{k})$ for $n \in \{N_{k - 1}^{(j)} + 1, \dots, N_{k}^{(j)}\}$. Furthermore, $\beta = \beta(\V{x}_k)$, $\alpha_n = \alpha(\V{y}_{k, n}^{(j)})$, $\xi = \xi(\eta_{k}^{(j)})$, $\iota = \iota(\V{x}_k ; \V{z}_k^{(j)})$, $\kappa_n = \kappa(\V{y}_{k, n}^{(j)}; \V{z}_k^{(j)})$, and $\nu = \nu(\eta_{k}^{(j)}; \V{z}_k^{(j)})$.}
    \label{fig:factor_graph}
    \vspace{-5.5mm}
\end{figure*}

\section{The Proposed BP Algorithm} \label{sec:bp}
\vspace{-.5mm}

In this section, we derive the proposed \ac{bp} method for multipath-based \ac{slam}. \ac{bp} is an efficient approach for solving high-dimension Bayesian inference problems. A BP method is derived by performing local operations called ``messages'' on the edges of the factor graph \cite{KscFreLoe:01,YedFreWei:05,KolFri:B09} that represents the statistical model of the Bayesian estimation problem. When the factor graph is a tree, the solutions provided the \ac{bp}, called ``beliefs'', are equal to the true marginal posterior \acp{pdf} required for the computation of optimum estimates. However, the factor graph representing the statistical model for direct SLAM shown in Fig.~\ref{fig:factor_graph} has loops. Thus, the beliefs computed by \ac{bp} are only approximations of the true marginal posteriors \cite{KscFreLoe:01,YedFreWei:05,KolFri:B09}. Despite the fact that theoretical performance guarantees are typically unavailable, performing BP on graphs with loops is a widely used strategy to compute accurate approximations of marginal posterior \acp{pdf} with application in, e.g., modern channel coding \cite{RicUrb:B08}, cooperative localization \cite{WymLieWin:09}, mulitobject tracking \cite{MeyKroWilLauHlaBraWin:J18}, and \ac{slam} \cite{LeiMeyHlaWitTufWin:J19}. In graphs with loops, there is some flexibility regarding the order in which messages are computed. Different message-passing orders may lead to different beliefs. In this work, we make use of the following message passing computation order: (i) messages are passed only forward in time, (ii) along the edges connecting an agent state variable node ``$\V{x}_{k}$'' and a new \ac{pf} state variable node ``$\V{y}_{k, n}^{(j)}$'', messages are only sent from the former to the latter \cite{LeiMeyHlaWitTufWin:J19}, (iii) all messages are only computed once, i.e., messages are not updated iteratively. \ac{bp} messages are also shown in Fig.~\ref{fig:factor_graph}. They are computed in parallel for all \acp{pa} $j \in \{1, \dots, J\}$. All messages developed in what follows are obtained from conventional BP message rules for messages passed from factor to variable nodes, and messages passed from variable nodes to factor nodes \cite{KscFreLoe:01,YedFreWei:05,KolFri:B09}\vspace{-3mm}.


\subsection{Prediction and Birth Messages} \label{subsec:bp_pred}
\vspace{-1mm}

First, we present the prediction message that is passed from factor node ``$f(\V{x}_k | \V{x}_{k - 1})$'' to variable node ``$\V{x}_k$''. This message is obtained\vspace{-.7mm} as
\begin{equation}
    \beta(\V{x}_k) = \int f(\V{x}_k | \V{x}_{k - 1}) \tilde{f}(\V{x}_{k - 1}) \hspace{1mm} \mathrm{d} \V{x}_{k - 1} \label{eq:bp_beta}
    \vspace{-.7mm}
\end{equation}
where $\tilde{f}(\V{x}_{k - 1})$ is the belief of the agent state at the previous time step $k - 1$.

Next, we develop the messages that are sent from the factor nodes ``$f(\V{y}_{k, n}^{(j)} | \V{y}_{k - 1, n}^{(j)})$'', $n \in \{1, \dots, N_{k}^{(j)}\}$, $j \in \{1, \dots, J\}$ to the corresponding \ac{pf} variable nodes ``$\V{y}_{k, n}^{(j)}$''. For each legacy \ac{pf} state\vspace{-.5mm} $\V{y}_{k, n}^{(j)}$ with $n \in \{1, \dots, N_{k - 1}^{(j)}\}$, $j \in \{1, \dots, J\}$, this message is given\vspace{-1mm} by
\begin{align}
	\alpha(\V{y}_{k, n}^{(j)}) = \sum_{r_{k - 1, n}^{(j)} \in \{0, 1\} } &\int f(\V{\phi}_{k, n}^{(j)}, r_{k, n}^{(j)} | \V{\phi}_{k - 1, n}^{(j)}, r_{k - 1, n}^{(j)}) \nn \\[-1mm]
	&\times \tilde{f}(\V{\phi}_{k - 1, n}^{(j)}, r_{k - 1, n}^{(j)}) \hspace{1mm} \mathrm{d} \V{\phi}_{k - 1, n}^{(j)} \label{eq:bp_alpha_legacy} \\[-5mm]
	\nn
\end{align}
where $\tilde{f}(\V{y}_{k - 1, n}^{(j)}) = \tilde{f}(\V{\phi}_{k - 1, n}^{(j)}, r_{k - 1, n}^{(j)})$ is the corresponding belief at time step $k - 1$. Specifically, by inserting \eqref{eq:state_transition_pf1} and \eqref{eq:state_transition_pf2} into \eqref{eq:bp_alpha_legacy}, for $r_{k - 1, n}^{(j)} = 1$, we obtain
\begin{equation}
    \alpha(\V{\phi}_{k, n}^{(j)}, 1) = \int p_{\mathrm{s}} f(\V{\phi}_{k, n}^{(j)} | \V{\phi}_{k - 1, n}^{(j)}) \tilde{f}(\V{\phi}_{k - 1, n}^{(j)}, 1) \hspace{1mm} \mathrm{d} \V{\phi}_{k - 1, n}^{(j)}. \nn
\end{equation}
Similarly, for $r_{k - 1, n}^{(j)} = 0$ we have $\alpha(\V{\phi}_{k, n}^{(j)}, 0) = \alpha_{k, n}^{(j)} f_{\mathrm{D}}(\V{\phi}_{k, n}^{(j)})$ with
\begin{equation}
    \alpha_{k, n}^{(j)} \triangleq \int (1 - p_{\mathrm{s}}) \tilde{f}(\V{\phi}_{k - 1, n}^{(j)}, 1) + \tilde{f}(\V{\phi}_{k - 1, n}^{(j)}, 0) \hspace{1mm} \mathrm{d} \V{\phi}_{k - 1, n}^{(j)}. \nn
    \vspace{-1mm}
\end{equation}

\ac{pf} states $\V{y}_{k, n}^{(j)}$, $n \rmv\in\rmv \{N_{k - 1}^{(j)} + 1, \dots, N_{k}^{(j)}\}$, $j \rmv\in\rmv \{1, \dots, J\}$ are newly introduced at time $k$. The message sent from factor node ``$f(\V{y}_{k, n}^{(j)} | \V{x}_{k })$'' to variable node ``$\V{y}_{k, n}^{(j)}$'' is given\vspace{-1.5mm} by
\begin{equation}
    \alpha(\V{y}_{k, n}^{(j)}) = \int f(\V{y}_{k, n}^{(j)} | \V{x}_{k}) \beta(\V{x}_{k}) \hspace{1mm} \mathrm{d} \V{x}_{k}. \label{eq:bp_alpha_new}
\end{equation}
This message provides feature birth information. By inserting \eqref{eq:birth_pf} into \eqref{eq:bp_alpha_new} we obtain
\begin{equation}
    \alpha(\V{\phi}_{k, n}^{(j)}, 1) = p_{\mathrm{B}, n}^{(j)} \int f_{\mathrm{B}, n}^{(j)}(\V{\phi}_{k, n}^{(j)} | \V{x}_{k}) \beta(\V{x}_k) \hspace{1mm} \mathrm{d} \V{x}_k \nn
\end{equation}
and $\alpha(\V{\phi}_{k, n}^{(j)}, 0) = (1 - p_{\mathrm{B}, n}^{(j)}) f_{\mathrm{D}}(\V{\phi}_{k, n}^{(j)})$. 

Finally, for each $j \rmv\in\rmv \{1, \dots, J\}$, the prediction message that is sent from the factor node ``$f(\eta_{k}^{(j)} | \eta_{k - 1}^{(j)})$'' to the noise variance node ``$\eta_{k}^{(j)}$'', is given by
\begin{equation}
    \xi(\eta_{k}^{(j)}) = \int \hspace{-.5mm} f(\eta_{k}^{(j)} | \eta_{k - 1}^{(j)}) \ist \tilde{f}(\eta_{k - 1}^{(j)}) \hspace{.5mm} \mathrm{d} \eta_{k - 1}^{(j)}. \label{eq:bp_xi}
\end{equation}
where $\tilde{f}(\eta_{k - 1}^{(j)})$ is the belief of the noise variance computed at the previous time step $k - 1$. 

Following \ac{bp} message passing rules, messages sent from the \ac{pf} state nodes ``$\V{y}_{k, n}^{(j)}$,'' $n \rmv\in\rmv \{1, \dots, N_{k}^{(j)}\}$, $j \rmv\in\rmv \{1, \dots, J\}$ to the factor nodes ``$f(\V{z}_{k}^{(j)} | \V{x}_k, \V{y}_{k}^{(j)}, \eta_{k}^{(j)})$'' are equal to $\alpha(\V{y}_{k, n}^{(j)})$, while messages sent from the noise variance nodes ``$\eta_{k}^{(j)}$,'' $j \rmv\in\rmv \{1, \dots, J\}$ to ``$f(\V{z}_{k}^{(j)} | \V{x}_k, \V{y}_{k}^{(j)}, \eta_{k}^{(j)})$'' are equal to $\xi(\eta_{k}^{(j)})$. In addition, due to our message schedule stated above, no message is passed from ``$f(\V{y}_{k, n}^{(j)} | \V{x}_k)$'', $n \in \{N_{k - 1}^{(j)} + 1, \dots, N_k^{(j)}\}$, $j \rmv\in\rmv \{1, \dots, J\}$  to ``$\V{x}_{k}$.'' Thus, the messages sent from ``$\V{x}_{k}$'' to factor nodes ``$f(\V{y}_{k, n}^{(j)} | \V{x}_k)$,'' $n \in \{N_{k - 1}^{(j)} + 1, \dots, N_k^{(j)}\}$, $j \rmv\in\rmv \{1, \dots, J\}$ and ``$f(\V{z}_{k}^{(j)} | \V{x}_k, \V{y}_{k}^{(j)}, \eta_{k}^{(j)})$,''  $j \rmv\in\rmv \{1, \dots, J\}$ are equal to $\beta(\V{x}_k)$. It is worth mentioning that the messages resulting from the equations \eqref{eq:bp_beta}-\eqref{eq:bp_xi} are all \acp{pdf}, i.e., they all normalized to one. This is because the messages needed as an input in \eqref{eq:bp_beta}-\eqref{eq:bp_xi}, namely $ \tilde{f}(\V{x}_{k - 1})$, $\tilde{f}(\eta_{k}^{(j)})$, $ \beta(\V{x}_{k}) $, and $ \tilde{f}(\eta_{k - 1}^{(j)})$ are also  \acp{pdf}. Message computations in  \eqref{eq:bp_beta}-\eqref{eq:bp_xi} can then be seen as marginalization from joint distributions that consist of two\vspace{-4mm} variables.


\subsection{Measurement Update Messages} \label{subsec:bp_meas}

After prediction and birth messages are obtained, a measurement update is performed where messages that provide the information of measurements $\V{z}_{k}$ are obtained. Specifically, messages sent from factor nodes ``$f(\V{z}_{k}^{(j)} | \V{x}_k, \V{y}_{k}^{(j)}, \eta_{k}^{(j)})$'' to agent state ``$\V{x}_k$'', \ac{pf} states ``$\V{y}_{k, n}^{(j)}$,'' $n \rmv\in\rmv \{1, \dots, N_{k}^{(j)}\}$, $j \rmv\in\rmv \{1, \dots, J\}$  and noise variance ``$\eta_{k}^{(j)}$,'' $j \rmv\in\rmv \{1, \dots, J\}$ are computed. In particular, the messages $\iota(\V{x}_k; \V{z}_k^{(j)} )$, $j \rmv\in\rmv \{1, \dots, J\}$ sent from ``$f(\V{z}_{k}^{(j)} | \V{x}_k, \V{y}_{k}^{(j)}, \eta_{k}^{(j)})$'' to ``$\V{x}_k$'' are obtained\vspace{-1mm} as
\begin{align}
    \iota(\V{x}_k; \V{z}_k^{(j)} ) =& \sum_{r_{k, 1}^{(j)} \in \{0, 1\}}  \hspace{1mm} \cdots \hspace{-1mm} \sum_{r_{k, N_k^{(j)}}^{(j)} \in \{0, 1\}} \nn \\[-.5mm]
    & \int \cdots \int f(\V{z}_{k}^{(j)} | \V{x}_k, \V{y}_{k}^{(j)}, \eta_{k}^{(j)}) \xi(\eta_{k}^{(j)})  \nn \\[.5mm]
    & \times \prod_{n = 1}^{N_k^{(j)}} \alpha(\V{y}^{(j)}_{k, n}) \hspace{1mm} \mathrm{d} \V{\phi}_{k, 1}^{(j)} \cdots \mathrm{d} \V{\phi}_{k, N_k^{(j)}}^{(j)} \hspace{1mm} \mathrm{d} \eta_{k}^{(j)}. \label{eq:bp_iota} \\[-6.5mm]
    \nn
\end{align}
Note that the measurements $\V{z}_k^{(j)}$, $j \rmv\in\rmv \{1, \dots, J\}$ are observed and thus fixed. This is indicated by the ``;'' notation in $\iota(\V{x}_k; \V{z}_k^{(j)} )$. These messages provide new agent location information related to measurement $\V{z}_k^{(j)}$ by leveraging knowledge on \acp{pf} states and the noise covariance matrix.


Next, the messages $\kappa(\V{y}_{k, n}^{(j)}; \V{z}_{k}^{(j)} )$,  $n \in \{1, \dots, N_{k}^{(j)}\}$, $j \rmv\in\rmv \{1, \dots, J\}$ passed from ``$f(\V{z}_{k}^{(j)} | \V{x}_k, \V{y}_{k}^{(j)}, \eta_{k}^{(j)})$'' to \ac{pf} ``$\V{y}_{k, n}^{(j)}$'' are given\vspace{0mm} by
\begin{align}
    \hspace{-10mm}& \kappa(\V{y}_{k, n}^{(j)}; \V{z}_{k}^{(j)} ) = \nn \\[2mm]
    & \sum_{r_{k, 1}^{(j)} \in \{0, 1\}} \cdots \sum_{r_{k, n - 1}^{(j)} \in \{0, 1\}} \sum_{r_{k, n + 1}^{(j)} \in \{0, 1\}} \cdots \sum_{r_{k, N_k^{(j)}}^{(j)} \in \{0, 1\}} \nn \\[0mm]
    & \int \cdots \int f(\V{z}_k^{(j)} | \V{x}_k, \V{y}_k^{(j)}, \eta_{k}^{(j)}) \xi(\eta_{k}^{(j)}) \beta(\V{x}_k)  \prod_{\substack{n' = 1 \\ n' \ne n}}^{N_k^{(j)}} \alpha(\V{y}^{(j)}_{k, n'}) \nn \\[2.5mm]
    & \mathrm{d} \V{\phi}_{k, 1}^{(j)} \cdots \mathrm{d} \V{\phi}_{k, n - 1}^{(j)} \hspace{1mm} \mathrm{d} \V{\phi}_{k, n + 1}^{(j)} \cdots \mathrm{d} \V{\phi}_{k, N_k^{(j)}}^{(j)} \hspace{1mm} \mathrm{d} \eta_{k}^{(j)} \hspace{1mm} \mathrm{d} \V{x}_k. \label{eq:bp_kappa} \\[-4.5mm]
    \nonumber
\end{align}
These messages provide new location information related to measurement $\V{z}_k^{(j)}$ for each \ac{pf} state by taking the knowledge on all the other \acp{pf} states, on the agent state, and on the noise covariance matrix into account.

Finally, the messages $\nu(\eta_{k}^{(j)}; \V{z}_k^{(j)} )$, $j \rmv\in\rmv \{1, \dots, J\}$ sent from ``$f(\V{z}_{k}^{(j)} | \V{x}_k, \V{y}_{k}^{(j)}, \eta_{k}^{(j)})$'' to the noise variance nodes ``$\eta_{k}^{(j)}$'' are calculated\vspace{-.5mm} as
\begin{align}
    \nu(\eta_{k}^{(j)}; \V{z}_k^{(j)} ) =& \sum_{r_{k, 1}^{(j)} \in \{0, 1\}} \cdots \sum_{r_{k, N_k^{(j)}}^{(j)} \in \{0, 1\}} \nn \\[-2mm]
    & \int \cdots \int f(\V{z}_{k}^{(j)} | \V{x}_k, \V{y}_{k}^{(j)}, \eta_{k}^{(j)}) \prod_{n = 1}^{N_k^{(j)}} \alpha(\V{y}^{(j)}_{k, n})  \nn \\[2.5mm]
    & \times \beta(\V{x}_k) \hspace{1mm} \mathrm{d} \V{\phi}_{k, 1}^{(j)} \cdots \mathrm{d} \V{\phi}_{k, N_k^{(j)}}^{(j)} \hspace{1mm} \mathrm{d} \V{x}_k. \label{eq:bp_nu}
\end{align}
These messages provide new noise variance information related to measurement $\V{z}_k^{(j)}$ by taking the knowledge on all the \acp{pf} states and on the agent state into\vspace{-4mm} account.

\subsection{Approximate Update and Belief Calculation}
\vspace{-.5mm}
\label{subsec:bp_approx}

The computation of measurement update messages in \eqref{eq:bp_iota}--\eqref{eq:bp_nu} involves marginalization of $N_k^{(j)}$ or $N_k^{(j)} \rmv-\rmv 1$ \ac{pf} states. The complexity of this marginalization scales exponentially with the number of \acp{pf} $N_k^{(j)}$. To improve the scalability of the proposed algorithm, we perform the following approximations. Since the likelihood function $f(\V{z}_{k}^{(j)} | \V{x}_k, \V{y}_{k}^{(j)}, \eta_{k}^{(j)})$ is zero-mean complex Gaussian with respect to $\V{z}_k^{(j)}$, the messages computed in \eqref{eq:bp_iota}--\eqref{eq:bp_nu}, i.e., $\iota(\V{x}_k; \V{z}_k^{(j)} )$, $\kappa(\V{y}_{k, n}^{(j)}; \V{z}_{k}^{(j)} )$, and $\nu(\eta_{k}^{(j)}; \V{z}_k^{(j)} )$, can be interpreted as zero-mean complex Gaussian mixtures \acp{pdf} of $\V{z}_k^{(j)}$. We here approximate them with zero-mean Gaussian \acp{pdf} via moment matching \cite{DavGar:22,LiaKroMey:J23}, i.e., we compute the following approximated messages
$\tilde{\iota} (\V{x}_k; \V{z}_k^{(j)} ) = \mathcal{CN}\big(\V{z}_k^{(j)}; \V{0}, \M{C}_{\iota, k}^{(j)}(\V{x}_k)\big),$
$\tilde{\kappa} (\V{y}_{k, n}^{(j)}; \V{z}_{k}^{(j)} ) = \mathcal{CN}\big(\V{z}_k^{(j)}; \V{0}, \M{C}_{\kappa, k, n}^{(j)}\big(\V{y}^{(j)}_{k,n}\big) \big)$, and $
\tilde{\nu} (\eta_{k}^{(j)}; \V{z}_k^{(j)} )$ $= \mathcal{CN}(\V{z}_k^{(j)}; \V{0}, \M{C}_{\nu, k}^{(j)})$
such that $\M{C}_{\iota, k}^{(j)}(\V{x}_k)$, $\M{C}_{\kappa, k, n}^{(j)}\big(\V{y}^{(j)}_{k,n}\big)$, and $\M{C}_{\nu, k}^{(j)}$ are the covariance of the original Gaussian mixtures $\iota(\V{x}_k; \V{z}_k^{(j)} )$, $\kappa(\V{y}_{k, n}^{(j)}; \V{z}_{k}^{(j)} )$, and $\nu(\eta_{k}^{(j)}; \V{z}_k^{(j)} )$, respectively. In particular, the matrices $\M{C}_{\iota, k}^{(j)} (\V{x}_k)$, $ \M{C}_{\kappa, k, n}^{(j)}\big(\V{y}^{(j)}_{k,n}\big)$, and $\M{C}_{\nu, k}^{(j)}$ are obtained as\vspace{-2mm} \cite{LiaLeiMey:J24SM}
\begin{align}
    \M{C}_{\iota, k}^{(j)}(\V{x}_k) &= \sum_{n = 1}^{N_k^{(j)}} \M{C}_{1, k, n}^{(j)} \rmv (\V{x}_k)  + \eta_{\xi, k}^{(j)} \M{I}_M \label{eq:bp_cov_iota} \\[-.7mm]
    \hspace{-3mm} \M{C}_{\kappa, k, n}^{(j)} \big(\V{y}^{(j)}_{k,n}\big)  &= r_{k, n}^{(j)} \M{C}_{2, k, n}^{(j)} \big(\V{\phi}^{(j)}_{k,n}\big) + \sum_{\substack{n' = 1 \\ n' \ne n}}^{N_k^{(j)}} \M{C}_{3, k, n'}^{(j)} + \eta_{\xi, k}^{(j)} \M{I}_M \nn\\[-5mm]
    \label{eq:bp_cov_kappa} \\[-1.7mm]
    \M{C}_{\nu, k}^{(j)} &= \eta_{k}^{(j)} \M{I}_M + \sum_{n = 1}^{N_k^{(j)}} \M{C}_{3, k, n}^{(j)} \label{eq:bp_cov_nu}
\end{align}
where we introduced the following matrices that have to be computed in a preparatory step, i.e.\vspace{1mm},
\begin{align}
\M{H}_{k, n}^{(j)}\rmv(\V{x}_k, \V{\phi}_{k, n}^{(j)}) &= \gamma_{k, n}^{(j)} \V{h}_{k, n}^{(j)} \V{h}_{k, n}^{(j) \CH} \nn \\[2mm]
    \M{C}_{1, k, n}^{(j)}(\V{x}_k) &= \int \M{H}_{k, n}^{(j)}\rmv(\V{x}_k, \V{\phi}_{k, n}^{(j)}) \ist \alpha(\V{\phi}_{k, n}^{(j)}, 1) \hspace{.3mm} \mathrm{d} \V{\phi}_{k, n}^{(j)} \label{eq:covUpdate1} \\[1mm]
    \M{C}_{2, k, n}^{(j)} \big(\V{\phi}^{(j)}_{k,n}\big) &= \int \M{H}_{k, n}^{(j)}\rmv(\V{x}_k, \V{\phi}_{k, n}^{(j)}) \ist \beta(\V{x}_{k}) \mathrm{d}\V{x}_{k}  \label{eq:covUpdate2} \\[2mm]
    \M{C}_{3, k, n}^{(j)} &= \int \M{H}_{k, n}^{(j)} \rmv (\V{x}_k, \V{\phi}_{k, n}^{(j)}) \ist \alpha(\V{\phi}_{k, n}^{(j)}, 1)  \nn\\[0mm]
    &\hspace{20mm}\times \beta(\V{x}_{k}) \hspace{.3mm} \mathrm{d} \V{\phi}_{k, n}^{(j)} \hspace{.3mm}  \mathrm{d} \V{x}_k \label{eq:covUpdate3} \\[.8mm]
    \eta_{\xi, k}^{(j)} &= \int \eta_{k}^{(j)} \ist \xi(\eta_{k}^{(j)}) \hspace{.3mm} \mathrm{d} \eta_{k}^{(j)}, \label{eq:covUpdate4}
\end{align}
Recall that $\V{h}_{k, n}^{(j)}$ is the contribution of the \ac{pf} with indexes $(j, n)$ as introduced in \eqref{eq:meas_model}. This contribution is a function of $\V{y}^{(j)}_{k,n}$ and $\V{x}_k$. Note that for $r^{(j)}_{k,n} \rmv=\rmv 0$, $\M{C}_{\kappa, k, n}^{(j)}\rmv\big(\V{y}^{(j)}_{k,n} \big) = \M{C}_{\kappa, k, n}^{(j)}\rmv\big(\V{\phi}^{(j)}_{k,n},r^{(j)}_{k,n} \big)$ in \eqref{eq:bp_cov_kappa} is independent of $\V{\phi}^{(j)}_{k,n}$. 


With the approximated \ac{bp} messages, the beliefs of agent state $\V{x}_k$, \ac{pf} states $\V{y}_{k, n}^{(j)}$, $n \in \{1, \dots, N_{k}^{(j)}\}$, $j \rmv\in\rmv \{1, \dots, J\}$ and noise variance $\eta_{k}^{(j)}$, $j \rmv\in\rmv \{1, \dots, J\}$ can be obtained as the product of all messages sent to the corresponding variable node\vspace{-4.5mm}, i.e.,
\begin{align}
    \tilde{f}(\V{x}_k) &\propto \beta(\V{x}_k) \prod^{J}_{j = 1} \tilde{\iota}(\V{x}_k; \V{z}_k^{(j)} ) \label{eq:belief1} \\[.7mm]
    \tilde{f}(\V{y}_{k, n}^{(j)}) &\propto \alpha(\V{y}_{k, n}^{(j)}) \ist \tilde{\kappa}(\V{y}_{k, n}^{(j)}; \V{z}_k^{(j)} ) \label{eq:belief2}  \\[2.5mm]
    \tilde{f}(\eta_{k}^{(j)}) &\propto  \xi(\eta_{k}^{(j)}) \ist \tilde{\nu}(\eta_{k}^{(j)}; \V{z}_k^{(j)} ). \label{eq:belief3} \\[-4mm]
    \nn
\end{align}
Computation of these beliefs involves proper normalization such that they integrate and sum to one. The calculated beliefs can then be used for declaration and state estimation as discussed in\vspace{-2mm} Section~\ref{subsec:declaration_estimation}.


%

\section{Particle-Based Implementation} \label{sec:particle}
\vspace{-0mm}

In general, it is not possible to calculate the beliefs and \ac{bp} messages discussed in Section~\ref{sec:bp} in closed form. This is because the integrations in \ac{bp} message passing equations \eqref{eq:bp_beta}--\eqref{eq:bp_xi} and  \eqref{eq:covUpdate1}--\eqref{eq:covUpdate4} as well as message multiplication in belief calculations in \eqref{eq:belief1}--\eqref{eq:belief3} cannot be performed analytically.

Hence, in this section, we present a computationally feasible particle-based implementation \cite{AruMasGorCla:02,IhlMca:09,MeyBraWilHla:J17} of the proposed \ac{bp} method. Here, the beliefs $\tilde{f}(\V{x}_k)$, $\tilde{f}(\V{y}_{k, n}^{(j)}) =  \tilde{f}(\V{\phi}_{k, n}^{(j)}, r_{k, n}^{(j)})$, and $\tilde{f}(\eta_{k}^{(j)})$ are represented by sets of weighted particles $\big\{(\V{x}_k^{(p)}, w_{\V{x}, k}^{(p)})\big\}_{p = 1}^P$, $\big\{(\V{\phi}_{k, n}^{(j, p)}, w_{\V{y}, k, n}^{(j, p)})\big\}_{p = 1}^P$, $n \in \{1, \dots, N_{k}^{(j)}\}$, $j \rmv\in\rmv \{1, \dots, J\}$ and $\big\{(\eta_{k}^{(j, p')}, w_{\eta, k}^{(j, p')})\big\}_{p' = 1}^{P'}$, $j \rmv\in\rmv \{1, \dots, J\}$. To keep the computational complexity linear in the number of particles, we adopted an approach introduced in \cite{MeyHliHla:J16} where the particles representing incoming BP messages for each agent-PF pair are stacked into a joint incoming message. This stacking approach requires that the agent state and the \ac{pf} states are represented by the same number of particles, $P$. The measurement noise variance can be represented by a different number of particles $P'$. Note that this stacking approach is asymptotically optimal \cite{MeyHliHla:J16}.
The weights representing PF states, $w_{\V{y}, k, n}^{(j, p)}$, do not sum to one, instead we have $p_{k, n}^{(j)} = \sum_{p = 1}^P w_{\V{y}, k, n}^{(j, p)} \approx \int \tilde{f}(\V{\phi}_{k, n}^{(j)}, 1) \hspace{1mm} \mathrm{d} \V{\phi}_{k, n}^{(j)}$, which approximately equals to the existence probability\vspace{-4.5mm} $f(r_{k, n}^{(j)} = 1 | \V{z}_{1 : k})$\vspace{-0mm}.

\subsection{Prediction and Birth} \label{sec:particle_pred}
\vspace{-1.5mm}

At each time step $k$, we have sets of weighted particles $\{(\V{x}_{k - 1}^{(p)}, w_{\V{x}, k - 1}^{(p)})\}_{p = 1}^P$, $\{(\V{\phi}_{k - 1, n}^{(j, p)}, w_{\V{y}, k - 1, n}^{(j, p)})\}_{p = 1}^P$, $n \in \{1, \dots, N_{k}^{(j)}\}$, $j \rmv\in\rmv \{1, \dots, J\}$ and $\{(\eta_{k - 1}^{(j, p')}, w_{\eta, k - 1}^{(j, p')})\}_{p' = 1}^{P'}$, $j \rmv\in\rmv \{1, \dots, J\}$ representing beliefs at the previous step $k \rmv-\rmv 1$. Since the messages $\beta(\V{x}_k)$, $\alpha(\V{y}_{k, n}^{(j)})$, and $\xi(\eta_{k}^{(j)})$ are \acp{pdf} as discussed in Section~\ref{subsec:bp_pred}, they can be represented with weighted particles $\{(\V{x}_{k}^{(p)}, w_{\beta, k}^{(p)})\}_{p = 1}^P$, $\{(\V{\phi}_{k, n}^{(j, p)}, w_{\alpha, k, n}^{(j, p)})\}_{p = 1}^P$, and $\{(\eta_{k}^{(j, p')}, w_{\xi, k}^{(j, p')})\}_{p' = 1}^{P'}$. These new sets of particles can be obtained as follows \cite{AruMasGorCla:02,IhlMca:09,MeyBraWilHla:J17}. 

For the agent state, one particle $\V{x}_k^{(p)}, p \in \{1, \dots, P\}$ is drawn from $f(\V{x}_k | \V{x}_{k - 1}^{(p)})$ for each $\V{x}_{k - 1}^{(p)}$. Similarly, for the noise covariance matrix, one particle $\eta_{k}^{(j, p')}, p' \in \{1, \dots, P'\}$ is drawn from $f(\eta_{k}^{(j)} | \eta_{k - 1}^{(j, p')})$ for each $\eta_{k - 1}^{(j, p')}$. Their weights remain unchanged, i.e., $w_{\beta, k}^{(p)} = w_{\V{x}, k - 1}^{(p)}$ and $w_{\xi, k}^{(j, p')} = w_{\eta, k - 1}^{(j, p')}$. For legacy \ac{pf} states, one particle $\V{\phi}_{k, n}^{(j, p)}, p \in \{1, \dots, P\},$ $n \in \{1, \dots, N_{k - 1}^{(j)}\}$ is drawn from $f(\V{\phi}_{k, n}^{(j)} | \V{\phi}_{k - 1, n}^{(j, p)})$ for each $\V{\phi}_{k - 1, n}^{(j, p)}$, and the particle weights is multiplied by the survival probability, i.e., $w_{\alpha, k, n}^{(j, p)} = p_{\mathrm{s}}w_{\V{y}, k - 1}^{(j, p)}$. The predicted existence probability of a PF state is thus $\sum_{p = 1}^P \rmv\rmv w_{\alpha, k, n}^{(j, p)} = p_{\mathrm{s}} \ist p_{k - 1, n}^{(j)}$. Finally,  for new \ac{pf} states $n \in \{ N_{k - 1}^{(j)} + 1, \dots, N_k^{(j)}  \} $, one particle $\V{\phi}_{k, n}^{(j, p)}, p \in \{1, \dots, P\}$ is drawn from $f_{\mathrm{B}, n}^{(j)}(\V{\phi}_{k, n}^{(j)} | \V{x}_{k}^{(p)})$ for each $\V{x}_{k}^{(p)}$, and the weights are set to\vspace{-5mm}  $w_{\alpha, k, n}^{(j, p)} = p_{\mathrm{B}, n}^{(j)} / P$\vspace{-0mm}.

\subsection{Measurement Update and Belief Calculation} \label{subsec:particle_meas}
\vspace{-1.7mm}

As discussed in Section~\ref{subsec:bp_approx}, we approximate the measurement update messages in \eqref{eq:bp_iota}--\eqref{eq:bp_nu} with zero-mean complex Gaussian \acp{pdf}. Thus, for each particle, we only need to compute approximations of the covariance matrices in \eqref{eq:bp_cov_iota}-\eqref{eq:bp_cov_nu} as\vspace{-3mm} follows
\begin{align}
    \tilde{\M{C}}_{\iota, k}^{(j)}\rmv\rmv\big(\V{x}^{(p)}_{k}\big) &= \sum_{n = 1}^{N_k^{(j)}} \tilde{\M{C}}_{1, k, n}^{(j)}\big(\V{x}^{(p)}_{k}\big)  \rmv+\rmv \tilde{\eta}_{\xi, k}^{(j)} \M{I}_M \nn\\[-9mm]
    \nn
    \end{align}
    \begin{align}
    \tilde{\M{C}}_{\kappa, k, n}^{(j)}\rmv\big(\V{y}^{(j,p)}_{k,n}\big) &= r_{k, n}^{(j)} \ist \tilde{\M{C}}_{2, k, n}^{(j)} \rmv\rmv\big(\V{\phi}^{(j,p)}_{k,n}\big) + \sum_{\substack{n' = 1 \\ n' \ne n}}^{N_k^{(j)}} \tilde{\M{C}}_{3, k, n'}^{(j)} \rmv+\rmv \tilde{\eta}_{\xi, k}^{(j)} \M{I}_M \nn \\[-1.5mm]
    \tilde{\M{C}}_{\nu, k}^{(j, p')} &= \eta_{k}^{(j, p')} \M{I}_M \rmv+\rmv \sum_{n = 1}^{N_k^{(j)}} \tilde{\M{C}}_{3, k, n}^{(j)}. \nn
\end{align}
Recall that for $r^{(j)}_{k,n} \rmv=\rmv 0$, $\tilde{\M{C}}_{\kappa, k, n}^{(j)}\rmv\big(\V{y}^{(j)}_{k,n} \big) = \tilde{\M{C}}_{\kappa, k, n}^{(j)}\rmv\big(\V{\phi}^{(j)}_{k,n},r^{(j)}_{k,n} \big)$ is independent of $\V{\phi}^{(j)}_{k,n}$. The resulting constant is denoted by $\tilde{\M{C}}_{\kappa, k, n}^{(j)} = \tilde{\M{C}}_{\kappa, k, n}^{(j)}\rmv\big(\V{\phi}^{(j,p)}_{k,n}\rmv\rmv,r^{(j)}_{k,n} \rmv=\rmv 0\big)$.

Here, the the covariance is \eqref{eq:covUpdate1}--\eqref{eq:covUpdate4} are performed by means on Monte Carlo\vspace{1mm} integration \cite{DouFreGor:01}
\begin{align} 
    \tilde{\M{H}}_{k, n}^{(j, p)} &= \gamma_{k, n}^{(j, p)} \V{h}_{k, n}^{(j, p)} \V{h}_{k, n}^{(j, p) \CH} \label{eq:particle_H} \\[2mm]
    \tilde{\M{C}}_{1, k, n}^{(j)} \big(\V{x}^{(p)}_{n}\big) &= \Big( \sum_{p' = 1}^P w_{\alpha, k, n}^{(j, p')} \Big) \tilde{\M{H}}_{k, n}^{(j, p)} \label{eq:particle_C1}  \\
    \tilde{\M{C}}_{2, k, n}^{(j)}\rmv\rmv\big(\V{\phi}^{(j,p)}_{k,n}\big) &=  \ist \tilde{\M{H}}_{k, n}^{(j, p)} \label{eq:particle_C2} \\[1mm]
    \tilde{\M{C}}_{3, k, n}^{(j)} &= \sum_{p = 1}^P w_{\beta, k}^{(p)} \ist \Big( \sum_{p' = 1}^P w_{\alpha, k, n}^{(j, p')} \Big) \tilde{\M{H}}_{k, n}^{(j, p)} \label{eq:particle_C3} \\
    \tilde{\eta}_{\xi, k}^{(j)} &= \sum_{p' = 1}^{P'} w_{\xi, k}^{(j, p')} \eta_{k}^{(j, p')} \label{eq:particle_Cnoise} 	
\end{align}
where $\V{h}_{k, n}^{(j, p)} = \V{h}(\tau_{k, n}^{(j, p)})$ is the contribution vector with delay $\tau_{k, n}^{(j, p)} = \Vert \V{p}_{k}^{(p)} -  \V{p}_{k, n}^{(j, p)} \Vert / c$ (cf.~Section~\ref{subsec:meas_model}) evaluated at the ``stacked'' particle $(\V{x}_k^{(p)}, \V{\phi}_{k, n}^{(j, p)})$. Details on how \eqref{eq:particle_H}--\eqref{eq:particle_Cnoise} are obtained based on the stacking approach \cite{MeyHliHla:J16} are provided in \cite{LiaLeiMey:J24SM}. We now compute a particle-based representation of \eqref{eq:belief1}--\eqref{eq:belief3} by means of importance sampling. In particular, for $p \in \{1,\dots,P\}$, we first calculate unnormalized weights\vspace{-.5mm} as
\begin{align} 
    \tilde{w}_{\V{x}, k}^{(p)}  &= w_{\beta, k}^{(p)} \prod_{j = 1}^J \mathcal{CN}\big(\V{z}_k^{(j)}; \V{0}, \tilde{\M{C}}_{\iota, k}^{(j)}\rmv\big(\V{x}^{(p)}_{k}\big)\big) \label{eq:weights1} \\[.5mm]
    \tilde{w}_{\V{y}, k, n}^{(j, p)} &= w_{\alpha, k, n}^{(j, p)} \mathcal{CN}\Big(\V{z}_k^{(j)}; \V{0}, \tilde{\M{C}}_{\kappa, k, n}^{(j)}\rmv\rmv\big(\V{\phi}^{(j,p)}_{k,n}\rmv\rmv,r^{(j)}_{k,n} \rmv=\rmv1 \big) \rmv \Big).  \label{eq:weights2}\\[-5mm]
    \nn
\end{align}
We also introduce the constant $\tilde{w}_{\V{y}, k, n}^{(j)} \rmv=\rmv  \mathcal{CN}\big(\V{z}_k^{(j)}; \V{0}, \tilde{\M{C}}_{\kappa, k, n}^{(j)} \big)$\vspace{.5mm}. Similarly, for $p' \in \{1,\dots,P'\}$, we obtain unnormalized weights as 
\begin{equation}
    \tilde{w}_{\eta, k}^{(j, p')} = w_{\xi, k}^{(j, p')} \mathcal{CN}\big(\V{z}_k^{(j)}; \V{0}, \tilde{\M{C}}_{\nu, k}^{(j, p')}\big) \label{eq:weights3}. 
\end{equation}
Here, we have replaced $\tilde{\iota}(\V{x}_k; \V{z}_k^{(j)} )$, $\tilde{\kappa}(\V{y}_{k, n}^{(j)}; \V{z}_k^{(j)} )$, and $\tilde{\nu}(\M{C}_{\V{\epsilon}, k}^{(j)}; \V{z}_k^{(j)} )$ in \eqref{eq:belief1}--\eqref{eq:belief3}, by $\mathcal{CN}\big(\V{z}_k^{(j)}; \V{0}, \tilde{\M{C}}_{\iota, k}^{(j)}\big(\V{x}^{(p)}_k\big)\big)$, $\mathcal{CN}\big(\V{z}_k^{(j)}; \V{0},$ $\tilde{\M{C}}_{\kappa, k, n}^{(j, p)}\big(\V{y}^{(j,p)}_{k,n}\big)\big)$, and $\mathcal{CN}\big(\V{z}_k^{(j)}; \V{0}, \tilde{\M{C}}_{\nu, k}^{(j, p')}\big)$, respectively. 

Next, the weighted particles $\{(\V{x}_{k}^{(p)}, w_{\V{x}, k}^{(p)})\}_{p = 1}^P$ and $\{(\eta_{k}^{(j, p')}, w_{\eta, k}^{(j, p')})\}_{p' = 1}^{P'}$ representing the beliefs $\tilde{f}(\V{x}_k)$ and $\tilde{f}(\eta_{k}^{(j)})$ are obtained by normalizing their corresponding weights, i.e., $w_{\V{x}, k}^{(p)} = \tilde{w}_{\V{x}, k}^{(p)} / \sum_{p = 1}^P \rmv \tilde{w}_{\V{x}, k}^{(p)}$ and $w_{\eta, k}^{(j, p')} = \tilde{w}_{\eta, k}^{(j, p')} / \sum_{p' = 1}^{P'} \rmv \tilde{w}_{\eta, k}^{(j, p')}$. For \ac{pf} states, the particle representations $\{(\V{\phi}_{k, n}^{(j, p)},$ $w_{\V{y}, k, n}^{(j, p)})\}_{p = 1}^P$ of the beliefs $\tilde{f}(\V{y}_{k, n}^{(j)}) =  \tilde{f}(\V{\phi}_{k, n}^{(j)}, r_{k, n}^{(j)})$ are obtained in a similar way. Here, the sum of weights represents the existence probability of the corresponding \ac{pf}, i.e., $p_{k, n}^{(j)} = \sum_{p = 1}^P w_{\V{y}, k, n}^{(j, p)} \approx \int \tilde{f}(\V{\phi}_{k, n}^{(j)}, 1) \hspace{1mm} \mathrm{d} \V{\phi}_{k, n}^{(j)}$. Properly normalized weights are obtained as (cf. \cite[Sec.~VI-C]{MeyBraWilHla:J17}\vspace{-1.5mm})
\begin{equation}
    w_{\V{y}, k, n}^{(j, p)} = \frac{\tilde{w}_{\V{y}, k, n}^{(j, p)}}{\sum_{p = 1}^P \tilde{w}_{\V{y}, k, n}^{(j, p)} + \tilde{w}_{\V{y}, k, n}^{(j)} \big( 1 - \sum_{p = 1}^P w_{\alpha, k, n}^{(j, p)} \big) }. \nn
    \vspace{1mm}
\end{equation}
Here, the denominator is a particle-based computation of the normalization constant\vspace{-1.8mm} $\sum_{r_{k, n}^{(j)} \in \{0, 1\}} \int \alpha(\V{\phi}_{k, n}^{(j)},$ $r_{k, n}^{(j)}) \ist \tilde{\kappa}(\V{\phi}_{k, n}^{(j)}, r_{k, n}^{(j)}; \V{z}_k^{(j)} ) \ist \mathrm{d} \V{\phi}_{k, n}^{(j)}$. To avoid particle degeneracy, a resampling step \cite{AruMasGorCla:02,IhlMca:09} is performed for the agent and each \ac{pf} at each time step.

Based on the probabilities $p_{k, n}^{(j)}$, $n \in \{1, \dots, N_{k}^{(j)}\}$, $j \rmv\in \{1, \dots, J\}$, the existence of \acp{pf} can now be determined. In particular, if $p_{k, n}^{(j)}$ is larger then the threshold $T_{\mathrm{dec}}$, \ac{pf} with indexes $(j, n)$ is declared to exist (cf. Section~\ref{subsec:declaration_estimation}). An approximation of the \ac{mmse} estimates in \eqref{eq:mmse_x}-\eqref{eq:mmse_y} at time $k$ for agent states and \acp{pf} declared to exist can now be computed as \vspace{-2.5mm}
\begin{align}
    \hat{\V{x}}_{k} = \sum_{p = 1}^P w_{\V{x}, k}^{(p)} \V{x}_{k}^{(p)} \hspace{5mm}
    \hat{\V{\phi}}_{k, n}^{(j)} = \frac{1}{p_{k, n}^{(j)}} \sum_{p = 1}^P w_{\V{y}, k, n}^{(j, p)} \V{\phi}_{k, n}^{(j, p)} \nn 
\end{align}
and the \ac{mmse} estimate of measurement noise variance in \eqref{eq:mmse_epsilon} can be approximated as $\hat{\eta}_{k}^{(j)} = \sum_{p' = 1}^{P'} w_{\eta, k}^{(j, p')} \eta_{k}^{(j, p')}$.

The number of \acp{pf} increases with each time step. To limit computational complexity, we prune \ac{pf} states that do not represent the \ac{pa} and have a low existence probability. In particular, if $p_{k, n}^{(j)}$ for $n \in \{2, \dots, N_k^{(j)}\}$ and $j \rmv\in\rmv \{1, \dots, J\}$ is lower than a certain threshold $T_{\mathrm{pru}}$, the corresponding \ac{pf} state is removed from the state\vspace{-3mm} space.

\subsection{Complexity Analysis} \label{subsec:particle_complexity}
\vspace{-.5mm}

The computational complexity of our particle-based implementation is analyzed in what follows. To simplify the notation, it is assumed that $P \rmv = \rmv P'$. The computation of prediction and birth messages only involves drawing particles and multiplying constants to weights for the agent, each \ac{pf}, and each measurement noise variance individually. It can easily be verified that the computational complexity of these operations scales as $\mathcal{O}\big(P \ist (N_k + J) \big)$, where $N_k = \sum_{j = 1}^{J} N_k^{(j)}$ is the total number of \acp{pf}. The computation of the measurement update messages and beliefs includes the covariance matrix calculations and the Gaussian \ac{pdf} evaluations. The complexity of the operations in \eqref{eq:particle_H}--\eqref{eq:particle_Cnoise} needed for the computation of covariance matrices is $\mathcal{O}(P N_k M^2)$. For complex Gaussian \ac{pdf} evaluations in \eqref{eq:weights1}--\eqref{eq:weights3}, the inverses of $\tilde{\M{C}}_{\iota, k}^{(j)}\big(\V{x}^{(p)}_k\big)$, $\tilde{\M{C}}_{\kappa, k, n}^{(j,p)}\big(\V{y}^{(p)}_{k,n}\big)$, and $\tilde{\M{C}}_{\nu, k}^{(j, p)}$ are needed for each particle and each \ac{pf}. The complexity of these operations scales as $\mathcal{O}(P N_k M^3)$. Thus, the overall complexity of the proposed particle-based \ac{bp} algorithm scales as $\mathcal{O}(P N_k M^3)$, i.e., linearly in the number of map features and cubically in the number of measurements. To further reduce computational complexity, we have developed an alternative approximate computation of \ac{bp} messages. The complexity of the resulting \ac{bp} message passing method scales quadraticly in the number of measurements $M$ and is thus particularly useful if the signal bandwidth is high. The alternative approximation and resulting BP messages are presented in \cite[Sec.~4]{LiaLeiMey:J24SM}. Experimental results show that direct SLAM with the alternative approximation performs almost identical to the original SLAM method but yields significantly lower runtime when the signal bandwidth is high \cite[Sec.~4]{LiaLeiMey:J24SM}\vspace{-4mm}.


%

\section{Numerical Results} \label{sec:exp}
\vspace{-.5mm}
In this section, to analyze the performance of the proposed direct SLAM method. We also present numerical results using synthetic and real data in \ac{2d} scenarios\vspace{-4mm}.

\begin{figure}[!t]
    \centering
    \hspace{-1mm}
    \resizebox{0.75\linewidth}{!}{\input{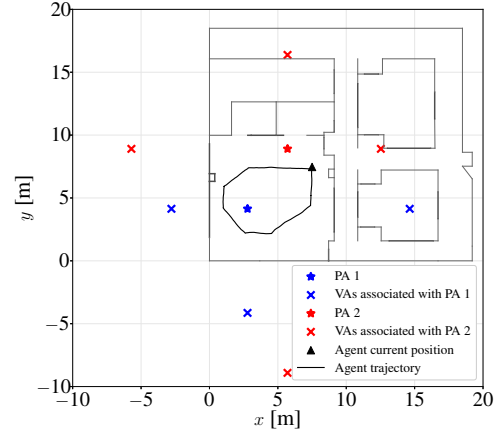}} \vspace{-2.5mm}
    \caption{The considered floor plan for performance evaluation. Stars indicate the PAs and the black line indicates the agent trajectory. The black triangle is the position of the agent at the current time step $k = 1$. The crosses are the first-order VAs and their colors indicate the PAs they associate with\vspace{-5mm}. }
    \label{fig:floorplan}
\end{figure}


\subsection{Experimental Setup} \label{subsec:exp_setup}
\vspace{-.5mm}

\subsubsection{System Model} \label{subsubsec:exp_setup_sys}
The agent state $\V{x}_k = [\V{p}_k^\T \hspace{1mm} \V{v}_k^\T]^\T \in \mathbb{R}^4$ consists of the agent's \ac{2d} position $\V{p}_k$ and \ac{2d} velocity $\V{v}_k$. Its state-transition function $f(\V{x}_k | \V{x}_{k - 1})$ is obtained from a constant-velocity motion model \cite[Sec. 4]{ShaKirLi:B02}. 
A Gamma distribution is used for the dynamics of the measurement noise variance, i.e., $f(\eta_{k}^{(j)} | \eta_{k - 1}^{(j)} ) = \mathcal{G} ( \eta_{k}^{(j)}; c_{\eta}, \eta_{k - 1}^{(j)} / c_{\eta} )$, which is of mean $\eta_{k - 1}^{(j)}$\vspace{-.3mm} and variance $(\eta_{k - 1}^{(j)})^2 / c_{\eta}$.
The evolution of the \ac{pf} position $\V{p}_{k, n}^{(j)}$ and intensity $\gamma_{k, n}^{(j)}$ are assumed to be independent, i.e., $f(\V{\phi}_{k, n}^{(j)} | \V{\phi}_{k - 1, n}^{(j)}) = f(\V{p}_{k, n}^{(j)} | \V{p}_{k - 1, n}^{(j)}) f(\gamma_{k, n}^{(j)} | \gamma_{k - 1, n}^{(j)})$. \acp{pf} are assumed static, i.e., the state-transition of \ac{pf} position would be a Dirac delta function. However, for the sake of robustness of our particle-based implementation, we use the random walk model with a small driving noise, i.e., $f(\V{p}_{k, n}^{(j)} | \V{p}_{k - 1, n}^{(j)}) = \mathcal{N}(\V{p}_{k, n}^{(j)}; \V{p}_{k - 1, n}^{(j)}, \sigma_{\V{p}, n}^2 \M{I}_2)$. The state transition of the intensity is modeled by a random walk model $f(\gamma_{k, n}^{(j)} | \gamma_{k - 1, n}^{(j)}) = \mathcal{N}(\gamma_{k, n}^{(j)} ; \gamma_{k - 1, n}^{(j)}, \sigma_{\gamma}^2)$. For new \acp{pf}, the birth of position and intensity are also assumed to be independent, i.e., $f_{\mathrm{B}, n}^{(j)}(\V{\phi} | \V{x}_k) = f_{\mathrm{B}, n}^{(j)}(\V{p} | \V{x}_k) f_{\mathrm{B}, n}^{(j)}(\gamma | \V{x}_k)$, where $f_{\mathrm{B}, n}^{(j)}(\V{p} | \V{x}_k)$ is the uniform \ac{pdf} over $\Set{P}_m(\V{x}_k)$ (cf. \eqref{eq:birth_pos_set}) and $f_{\mathrm{B}, n}^{(j)}(\gamma | \V{x}_k) = f_{\mathrm{B}, n}^{(j)}(\gamma)$ is the uniform \ac{pdf} over $[0, 2]$. 


\subsubsection{Parameters and Implementation} \label{subsubsec:exp_setup_parameter}
We consider an indoor environment with two \acp{pa} and the floor plan is shown in Fig.~\ref{fig:floorplan}. There are 679 time steps in total and the agent track is depicted by the black line. The driving noise variance of the agent state is set to $\sigma_{\V{x}}^2 = 10^{-4} \,\mathrm{m}^2/\mathrm{s}^2$. The driving noise variance for legacy \ac{pf} positions $\sigma_{\V{p}, n}^2$ is set to $10^{-8} \,\mathrm{m}^2$ for \acp{pa}, i.e., $n = 1$, and is set to $9 \times 10^{-6} \,\mathrm{m}^2$ for other \acp{va}, i.e., $n \ge 2$. \acp{pa} have lower driving noise variance as their positions are known. The driving noise variance for legacy \acp{pf}' intensity is $\sigma_{\gamma}^2 = 10^{-4}$ and the survival probability in \eqref{eq:state_transition_pf2} is $p_{\mathrm{s}} = 0.999$. The birth probability in \eqref{eq:birth_pf} is set to $p_{\mathrm{B}, n}^{(j)} = 10^{-4}$ for all new \acp{pf}. The state-transition parameter for noise variance is $c_{\eta} = 10$. In addition, we set the declaration threshold to $T_{\mathrm{dec}} = 0.5$ and the pruning threshold to $T_{\mathrm{pru}} = 10^{-2}$.


For the particle-based implementation, we choose $P = 10^4$ particles for the agent state and \ac{pf} states, and $P' = 1000$ particles for measurement noise variance. The particles for the initial agent position, $\{\V{p}_0^{(p)}\}_{p = 1}^P$, are drawn uniformly from a disk with center $\V{p}_1$ and radius 0.5m, where $\V{p}_1$ is the true agent position at $k \rmv=\rmv 1$. The initial agent velocity particles $\{\V{v}_0^{(p)}\}_{p = 1}^P$ are uniformly drawn from $[-0.01, 0.01]$m/s in both dimensions. The initial agent state particles are uniformly weighted, i.e., $w_{\V{x}, 0}^{(p)} = 1 / P$. The particles for initial \ac{pa} positions $\{\V{p}_{0, 1}^{(j, p)}\}_{p = 1}^P$ are sampled from a Gaussian \ac{pdf} $\mathcal{N}(\V{p}_{0, 1}^{(j, p)}; \V{p}_{1}^{(j)}, \sigma_{\V{p}, 1}^2 \M{I}_2)$ where 
$\V{p}_{1}^{(j)}$ is the position of \ac{pa} $j \in \{1, 2\}$. The particles for initial \ac{pa} intensity $\{\gamma_{0, 1}^{(j, p)}\}_{p = 1}^P$ are drawn uniformly from $[0, 2]$. There is no prior information for \acp{va}. Since the two \acp{pa} exist at the beginning for the considered agent track, we set the \ac{pa} state weights as $w_{\V{y}, 1}^{(j, p)} = 1 / P$. For the measurement noise variance, its initial particles $\{ \eta_{0}^{(j, p')} \}_{p' = 1}^{P'}$ are sampled uniformly from $0$ to $0.1$.


\subsubsection{Synthetic Data Generation} \label{subsubsec:exp_setup_synthetic} 

For synthetic data, only first-order \acp{va} are considered, i.e., every considered propagation path is reflected at most once from a feature in the environment. The parameters of paths are computed using a ray-tracing algorithm \cite{Bor:J84, Ant:B92}. The \acp{va} computed at $k = 1$ are shown in Fig.~\ref{fig:floorplan}, where we have $L_1^{(1)} = 4$ and $L_1^{(2)} = 5$ features corresponds to \ac{pa} 1 and 2, respectively. The ground truth \acp{va} are used exclusively for data generation and performance evaluation, and are not utilized in \ac{slam} algorithms. The transmitted signal spectrum $S(f)$ has a root-raised-cosine shape with a roll-off factor of $0$ (flat spectrum) and a bandwidth $B$ (specified in Sections~\ref{subsec:exp_synthetic} and \ref{subsec:exp_real}). We define the maximum observation distance as $d_\mathrm{max}=30\,$m, resulting in a frequency spacing of $\Delta = 10\,$MHz. The complex amplitude $\varrho_{k, l}^{(j)}$ are generated according to \eqref{eq:complex_amplitudes}, where the magnitude $a^{(j)}_{l,k} = 1$ for \acp{pa} ($l=1$) and $a^{(j)}_l = 0.7$ for \acp{va} ($l \ge 2$), which represents an attenuation of $1.5\,$dB for each reflection, and the phase of $a^{(j)}_{l,k}$ is drawn uniformly from $[0, 2\pi)$. The SNR output at $1\,$m distance to the mobile agent is set to be $42\,$dB resulting in $10^{-4.2}$ measurement noise variance.

To reduce computational complexity and increase robustness, instead of introducing all the $M$ new \acp{pf} at each time step, we here only establish a new \ac{pf} over a subset $\Set{N}_k^{(j)} \subseteq \{N_{k - 1}^{(j)} + 1, \dots, N_{k - 1}^{(j)} + M\}$. To do that, we first compute $\tilde{z}_{k, m}^{(j)} = \V{h}^{\CH}(\tau_m) \V{z}_k^{(j)}$ with $\tau_m = (m - 1)T_s$ and $\V{h}(t)$ introduced in \eqref{eq:ref_signal}. Note that the positive scalar values $|\tilde{z}_{k, m}^{(j)}|$, $m \in \{1, \dots, M\}$ represent the 1-D magnitude spectrum over the delay grid $[\tau_1 \cdots \tau_M]^\T$. A new \ac{pf} $n \in \Set{N}_k^{(j)}$, corresponding to $\tilde{z}_{k, m}^{(j)}, m = n - N_{k - 1}^{(j)}$ and occupying the region $\Set{P}_m(\V{x}_k)$ defined in \eqref{eq:birth_pos_set}, must satisfy the following two conditions: (i) the absolute value $|\tilde{z}_{k, m}^{(j)}|$ is larger than a certain threshold, $\gamma_{\mathrm{init}}^{(j)}$, and a local maximum with respect to $m$; (ii) the delay satisfies $(m - 1)T_s \cdot c < \min_p \Vert \V{p}_k^{(p)} - \V{p}_{k, 1}^{(j, p)} \Vert$ or $(m - 1)T_s \cdot c > \max_p \Vert \V{p}_k^{(p)} - \V{p}_{k, 1}^{(j, p)} \Vert$. Here, $\Vert \V{p}_k^{(p)} - \V{p}_{k, 1}^{(j, p)} \Vert$ is the Euclidean distance between the agent and \ac{pa} particles. The first condition ensures that only the peaks of the spectrum with values exceeding the threshold are selected. The second condition ensures that no PF with a delay similar to the PA is introduced. The threshold is set to $\gamma_{\mathrm{init}, k}^{(j)} = \sqrt{ \hat{\eta}_{k - 1}^{(j)} \cdot 10}$\vspace{-3mm}.

\subsection{Results of Synthetic Data} \label{subsec:exp_synthetic}
We perform 100 simulation runs for synthetic data, i.e., we generate 100 different measurements $\V{z}_k^{(j)}, j \in \{1, 2\}$ as discussed in Section~\ref{subsubsec:exp_setup_synthetic} and apply different methods independently. A bandwidth of $B = 300, 400$, and $600\,$MHz is considered. This results in a signal vector $\V{z}_k^{(j)}, j \in \{1, 2\}$ with $M=B/\Delta + 1 = 31, 41$, and $61$ samples, respectively. 


\begin{figure}[!t]
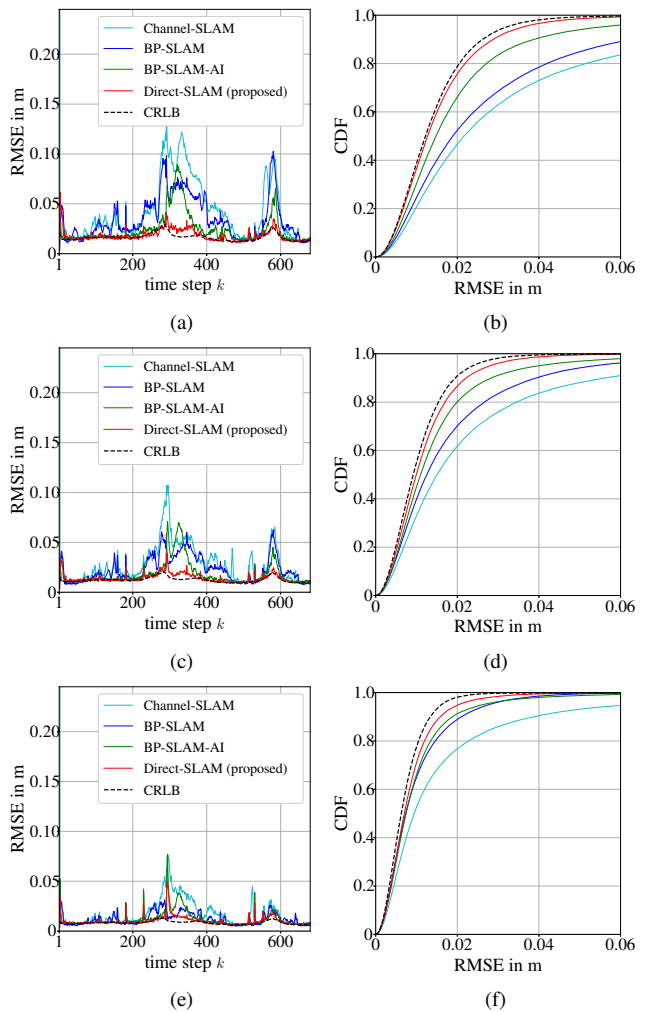

	\vspace{-3.5mm}
    \centering
    \subfloat[\hspace{-8mm} ]{\resizebox{0.48\linewidth}{!}{\input{Figs/rmse_bw300M_v2.pgf}} \label{fig:agent_synthetic_rmse_bw300M}} 
    \subfloat[\hspace{-4mm} ]{\resizebox{0.48\linewidth}{!}{\input{Figs/cdf_bw300M_v2.pgf}} \label{fig:agent_synthetic_cdf_bw300M}} \vspace{-2mm} \\
    \subfloat[\hspace{-8mm} ]{\resizebox{0.48\linewidth}{!}{\input{Figs/rmse_bw400M_v2.pgf}} \label{fig:agent_synthetic_rmse_bw400M}} 
    \subfloat[\hspace{-4mm} ]{\resizebox{0.48\linewidth}{!}{\input{Figs/cdf_bw400M_v2.pgf}} \label{fig:agent_synthetic_cdf_bw400M}} \vspace{-2mm} \\
    \subfloat[\hspace{-8mm} ]{\resizebox{0.48\linewidth}{!}{\input{Figs/rmse_bw600M_v2.pgf}} \label{fig:agent_synthetic_rmse_bw600M}} 
    \subfloat[\hspace{-4mm} ]{\resizebox{0.48\linewidth}{!}{\input{Figs/cdf_bw600M_v2.pgf}} \label{fig:agent_synthetic_cdf_bw600M}}
    \caption{Performance of agent localization on synthetic data: (a) Agent position RMSEs averaged over 100 simulation runs and (b) empirical CDFs of the RMSEs with $300\,$MHz bandwidth. (c)$\backslash$(d) and (e)$\backslash$(f) are the counterparts of (a)$\backslash$(b) for bandwidth $400\,$MHz and $600\,$MHz, respectively\vspace{-3mm}.}
    \label{fig:agent_synthetic}
\end{figure}

\begin{figure}[!t]
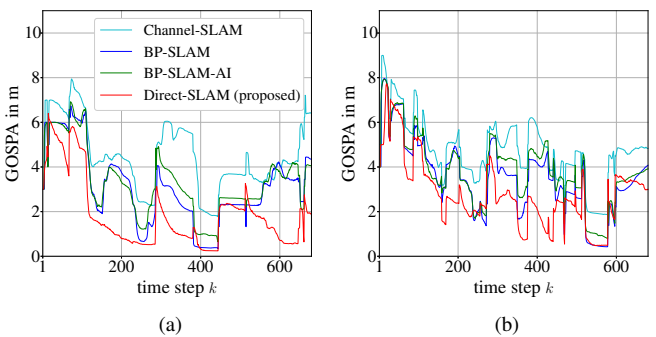

    \centering
    \subfloat[\hspace{-6mm} ]{\resizebox{0.48\linewidth}{!}{\input{Figs/gospa_sensor1_v2.pgf}}} \hspace{1mm}
    \subfloat[\hspace{-6mm} ]{\resizebox{0.48\linewidth}{!}{\input{Figs/gospa_sensor2_v2.pgf}}}
    \vspace{-1mm}
    \caption{Performance of mapping on synthetic data with 400MHz bandwidth: GOSPA error of estimated VAs associated with PA 1 (a)  and  PA 2 (b), averaged over 100 simulation runs.}
    \vspace{-3mm}
    \label{fig:gospa}
\end{figure}

\begin{figure}[!t]
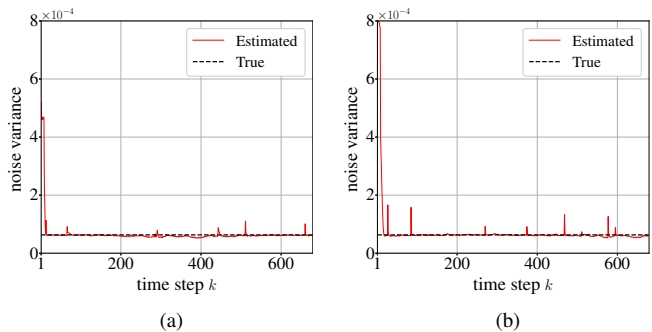

    \centering
    \subfloat[\hspace{-6mm} ]{\resizebox{0.48\linewidth}{!}{\input{Figs/noise_var_sensor1.pgf}}} \hspace{1mm}
    \subfloat[\hspace{-6mm} ]{\resizebox{0.48\linewidth}{!}{\input{Figs/noise_var_sensor2.pgf}}}
        \vspace{-2mm}
    \caption{True and estimated measurement noise variance $\eta_{k}^{(j)}$ averaged over 100 simulation runs for PA 1 (a)  and  PA 2 (b) at 400MHz bandwidth.}
    \vspace{-5mm}
    \label{fig:noise_cov}
\end{figure}

For performance evaluation, the following multipath-based \ac{slam} methods are used as benchmarks: ``Channel-SLAM'' \cite{GenJosWan:J16}, ``BP-SLAM'' \cite{LeiGreWit:19} and ``BP-SLAM-AI'' \cite{LeiGreLiTufWit:18}. All the reference methods use a sparse Bayesian learning channel estimator \cite{ShuWanJos:13,GerMecChrXenNan:J16,GreLeiWitFle:J24} to estimate \acp{mpc} from radio signals. The estimated \acp{mpc} are then fed to the downstream algorithm for \ac{slam}. Compared with BP-SLAM, in addition to using the delay information, BP-SLAM-AI further exploits the amplitude information in estimated \acp{mpc}. Fig.~\ref{fig:agent_synthetic} shows the \ac{rmse} of the estimated agent positions at different time steps averaged over all 100 simulation runs at different bandwidths, as well as the empirical \ac{cdf} of individual \acp{rmse} evaluated at all time steps and all simulation runs. The \ac{crlb} of the agent localization error \cite{TicMurNeh:J98, KalGeTalWymVal:21, LeiMeiRudDumWit:J15} is also included in Fig.~\ref{fig:agent_synthetic}. Note that, for the different bandwidths, in 3-5 runs of the total 100 runs, Channel-SLAM leads to an ``agent track loss''. An agent track loss is a run where the  \ac{rmse} rises above $1\,$m and continues to increase. Runs where the application of Channel-SLAM results in an agent track loss are not taken into account in the performance plots provided in Fig.~\ref{fig:agent_synthetic} (a), (c), and (e). In all 100 runs, the application of BP-SLAM, BP-SLAM-AI, and Direct-SLAM does not lead to an agent track loss. It can be seen that the proposed Direct-SLAM method closely follows the \ac{crlb} and outperforms the reference methods in terms of localization error. Especially, the performance gain is significant during time steps $k \in [100, 400]$ and around $k = 600$, where the geometry of the propagation environment is complicated, and many features have similar distances to the agent. Further analysis is provided \cite[Sec.~5.1]{LiaLeiMey:J24SM}. To quantify the mapping performance, we use the Euclidean distance-based \ac{gospa} metric \cite{RahGarSve:17} with cutoff parameter $c = 2\,$m and order $p = 1$. Results are shown in Fig.~\ref{fig:gospa}. The \ac{gospa} error of Direct-SLAM is generally smaller than the reference methods, and its performance gain is large in challenging situations, e.g., during $k \in [100, 400]$. Further qualitative results on mapping are shown in \cite{LiaLeiMey:J24SM}. Videos of this experiment that indicated a faster mapping performance of Direct-SLAM compared to BP-SLAM can be found in \cite{LiaLeiMey:J24SR}. Fig.~\ref{fig:noise_cov} compares the true and estimated measurement noise variance at all time steps. The estimated variance is large at the beginning as the particles are initialized over a large interval $[0, 0.1]$, then it quickly converges to the true value.



All experiments are conducted on a single core of Intel Xeon Gold 5222 CPU. The average computation time of different methods for each time step is reported in Table~\ref{tab:runtime}. Note the runtime of BP-SLAM and BP-SLAM-AI includes the $0.98$, $0.99$, and $0.99\,$s processing time of the fast variational sparse Bayesian learning channel estimator \cite{ShuWanJos:13} for $300$, $400$, and $600\,$MHz bandwidth, respectively. The runtime of Direct-SLAM is somewhat higher than the ones of the two reference methods. For $300\,$MHz, where Direct-SLAM yields the most significant performance gain, the runtime is only increased by $15.5\%$\vspace{-1mm}.

\begin{table}[!t]
	\centering
	\caption{Average runtime of different methods per time step}
	\label{tab:runtime}
	\begin{tabular}{c|ccc}
	\hline
	\multirow{2}{*}{Method}                         & \multicolumn{3}{c}{Time (s)}                                       \\ \cline{2-4} 
	                                                & \multicolumn{1}{c|}{300MHz} & \multicolumn{1}{c|}{400MHz} & 600MHz \\ \hline
	BP-SLAM \cite{LeiMeyHlaWitTufWin:J19}           & \multicolumn{1}{c|}{1.03}   & \multicolumn{1}{c|}{1.04}   & 1.04   \\
	BP-SLAM-AI \cite{LeiGreLiTufWit:18}             & \multicolumn{1}{c|}{1.06}   & \multicolumn{1}{c|}{1.06}   & 1.07   \\
	Direct-SLAM (proposed)                          & \multicolumn{1}{c|}{1.19}   & \multicolumn{1}{c|}{1.53}   & 2.58   \\ \hline
	\end{tabular}
	\vspace{-3mm}
\end{table}

\subsection{Results of Real Data} \label{subsec:exp_real}

The performance of the proposed method has been evaluated on real data collected in a classroom previously used in \cite{MeiLeiWit:14}. The floorplan shown in Fig.~\ref{fig:particle_results_semRoom}. The two crosses are the position of the two \acp{pa} and the yellow line is the agent trajectory. The trajectory consist of 220 agent positions with $0.05\,$m spacing along the trajectory. This signal was measured using an M-sequence correlative channel sounder with frequency range $3-10\,$GHz and antennas with an approximately uniform radiation pattern in the azimuth plane and zeros in the floor and ceiling directions. Within the measured band, the actual signal band was selected by a filter with rooted-raised-cosine impulse response with a roll-off factor of 0, a bandwidth of $B = 400\,$MHz, and a center frequency of $f_{\mathrm{c}} = 6\,$GHz, leading to measurements $\V{z}_k^{(j)}, j \in \{1, 2\}$ with $M = 41$.

For the experiments running on real data, we set the driving noise variance for legacy \ac{pf} positions $\sigma_{\V{p}, n}^2$ to $1 \times 10^{-6} \,\mathrm{m}^2$ for \acp{pa}, and to $6.4 \times 10^{-5} \,\mathrm{m}^2$ for other \acp{va}. The driving noise variance for legacy \acp{pf}' intensity is $\sigma_{\gamma}^2 = 10^{-2}$. In the likelihood function \eqref{eq:likelihood}, we artificially increase the measurement noise variance by a factor of 2. Note that, compared to the simulated scenario in Section \ref{subsec:exp_synthetic}, we increase the driving noise variance and the measurement noise variance to capture model mismatch, e.g., due to not explicitly modeling \ac{dmc}. In this way, we represent the mismatch between our statistical model and the true data-generating process as an additional noise source. The number of particles of the agent and \ac{pf} states is $P = 30000$ and the number of particles of noise variances is $P' = 3000$. The survival probability if $p_{\mathrm{s}} = 0.98$. The driving noise variance of the agent state is set to $\sigma_{\V{x}}^2 = 4 \times 10^{-4} \,\mathrm{m}^2/\mathrm{s}^2$. All other parameters remain the same.

For the reference method BP-SLAM and BP-SLAM-AI, $150000$ particles are used. In total, the data is processed in 30 runs, where only particle realizations are different across runs. The qualitative result of Direct-SLAM at a single time step in a single simulation run is shown in Fig.~\ref{fig:particle_results_semRoom}. The \ac{rmse} of the estimated agent positions at different time steps averaged over all 30 runs, as well as the empirical \ac{cdf} of individual \acp{rmse} evaluated at all time steps and all simulation runs are shown in Fig.~\ref{fig:agent_real}, and Direct-SLAM outperforms the reference methods with a clear margin. As the reference methods, the proposed method also does not consider non-specular propagation effects such as diffuse scattering and diffraction. Nonetheless, it can achieve robust performance in real scenarios\vspace{-5mm}.

\begin{figure}[!t]
    \centering
    \psfrag{x}[c][c][0.8]{$x$ [m]}
    \psfrag{y}[c][c][0.8]{$y$ [m]}
    \includegraphics[width=0.65\linewidth]{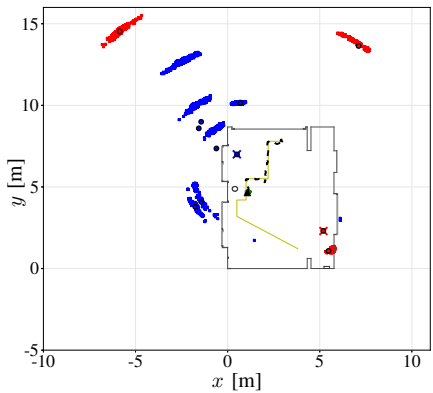}
    \vspace{-1.5mm}
    \caption{Results using real data for $k = 102$. Blue and red indicate features associated with the two PAs. Crosses depict the positions of PAs. Bullets are particles of declared PF positions, and circles with black edges are estimated positions of declared PFs. The yellow line is the expected agent trajectory, the black dashed line is the estimated agent trajectory, and the black triangle is the estimated agent position at the current time step\vspace{2mm}.}
    \vspace{-1.5mm}
    \label{fig:particle_results_semRoom}
\end{figure}

\begin{figure}[!t]
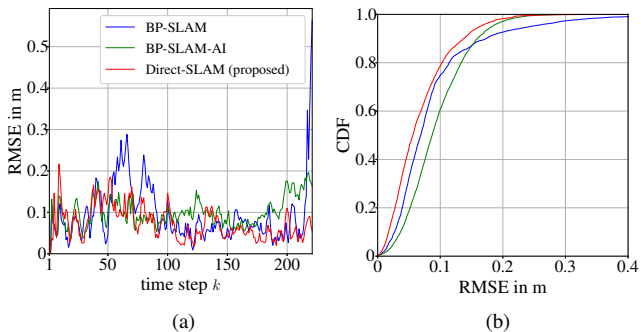

	\vspace{-3.5mm}
    \centering
    \subfloat[\hspace{-8mm} ]{\resizebox{0.48\linewidth}{!}{\input{Figs/rmse_semRoom.pgf}} \label{fig:agent_real_rmse_bw400M}} 
    \subfloat[\hspace{-4mm} ]{\resizebox{0.48\linewidth}{!}{\input{Figs/cdf_semRoom.pgf}} \label{fig:agent_real_cdf_bw400M}} 
    \vspace{-1.5mm}
    \caption{Performance of agent localization on real data: (a) Agent position RMSEs averaged over 30 simulation runs and (b) empirical CDFs of the RMSEs\vspace{2mm}.}
    \vspace{-5mm}
    \label{fig:agent_real}
\end{figure}

\acresetall
\section{Conclusion} \label{sec:conclusion}
We propose a novel \textit{Direct-SLAM} method for the multipath-based \ac{slam} problem. Our method aims to avoid the loss of information due to preprocessing by directly using the radio signal for \ac{slam} without using any channel estimator as a frontend. A new measurement model that describes the statistical relationship between the state of the mobile agent and map features is introduced. We also establish a birth model that makes it possible to infer the unknown and time-varying number of map features. A factor graph based on the new statistical model is established, and a \ac{bp} method that can efficiently compute approximate marginal \acp{pdf} (``beliefs'') for agent states and map features is derived. To reduce computational complexity, certain \ac{bp} messages are approximated by complex Gaussian \acp{pdf} via moment matching. Our complexity analysis shows that the proposed \ac{bp} method only scales linearly with the number of map features. Our numerical studies performed on synthetic data demonstrate that the proposed Direct-SLAM method achieves better localization and mapping accuracy compared with conventional methods that rely on a channel estimator. Performance gains are most significant when bandwidth is limited and in challenging environments when multiple propagation paths have similar lengths and are not resolvable by the channel estimator. Results on real data show that Direct-SLAM is robust even when non-specular \acp{mpc} is present. An interesting direction for future research is incorporating loop closer \cite{DisNewClaDurCso:01,MonThrKolWeg:02,CadCarCarLatScaNeiReiLeo:J16} and deep learning techniques \cite{SatWel:21,LiaMey:21,LiaMey:J23} into the current Direct-SLAM method. Another avenue for future work is to include non-specular propagation effects to make the measurement model more realistic, e.g., considering \ac{dmc} \cite{Ric:05}, and non-ideal reflecting surfaces \cite{WieVenWilLei:23,WieVenWilWitLei:J24} as well as introducing a state-transition-model to represent moving features\vspace{0mm}.






\ifCLASSOPTIONcaptionsoff
  \newpage
\fi


\renewcommand{\baselinestretch}{1.01}
\bibliographystyle{ieeetr}
\bibliography{IEEEabrv,StringDefinitions,Books,Papers,ref,refBooks}

\end{document}